\documentclass[useAMS,usenatbib]{mn2e}

\usepackage{amsmath}
\usepackage{natbib}
\usepackage{epsfig}
\usepackage{txfonts}
\usepackage{graphicx}
\usepackage{float}
\usepackage{bigints}

\usepackage{color,hyperref}
\definecolor{darkblue}{rgb}{0.0,0.0,0.5}
\hypersetup{colorlinks,breaklinks, linkcolor=darkblue,urlcolor=darkblue, anchorcolor=darkblue,citecolor=darkblue}

\def\ms{\,m\,s$^{-1}$}         %m.s -1
\def\kms{\,km\,s$^{-1}$}         %m.s -1
\def\ms{\hbox{\,m\,s$^{-1}$}}         %m.s -1
\def\cms{\hbox{\,cm\,s$^{-1}$}}       %cm.s -1
\def\m2s2{\hbox{\,m$^{2}$\,s$^{-2}$}} %m2.s -2
\def\kms{\hbox{\,km\,s$^{-1}$}}       %km.s -1
\def\vsini{\hbox{$\upsilon \sin i_{\star}\;$}}      %vsini
\def\Msun{\hbox{$\mathrm{M}_{\odot}$}}             %Msun

\def\Mjup{\hbox{$\mathrm{M}_{\rm Jup}$}}

\def\degr{\hbox{$^\circ$}}

\def\teff{T$_{\rm eff}$}
\def\logg{log {\it g}}
\def\met{[Fe/H]}

\def\vasy{V$_\mathrm{asy}$}
\def\vspan{V$_\mathrm{span}$}
\def\wspan{W$_\mathrm{span}$}

\def\aj{AJ}%
\def\apj{ApJ}%
\def\apjl{ApJ}%
\def\apjs{ApJS}%
\def\apss{Ap\&SS}%
\def\aap{A\&A}%
\def\aaps{A\&AS}%
\def\mnras{MNRAS}%
\def\pasp{PASP}%
\def\nat{Nature}%
\def\procspie{Proc.~SPIE}%
\title[PASTIS: Bayesian extrasolar planet validation]{PASTIS: Bayesian extrasolar planet validation \\
II. Constraining exoplanet blend scenarios using spectroscopic diagnoses}
\author[A. Santerne]{A. Santerne$^{1}$\thanks{E-mail:alexandre.santerne@astro.up.pt}, 
R.~F. D\'iaz$^{2}$, 
J.-M. Almenara$^{3}$,
F. Bouchy$^{4,2}$,
M. Deleuil$^{4}$,
P. Figueira$^{1}$,\newauthor
G. H\'ebrard$^{5,6}$,
C. Moutou$^{4,7}$,
S. Rodionov$^{4}$,
and N.~C. Santos$^{1,8}$
 \\
$^{1}$ Instituto de Astrof\'isica e Ci\^{e}ncias do Espa\c co, Universidade do Porto, CAUP, Rua das Estrelas, PT4150-762 Porto, Portugal\\
$^{2}$ Observatoire Astronomique de l'Universit\'e de Gen\`eve, 51 chemin des Maillettes, 1290 Versoix, Switzerland\\
$^{3}$ UJF-Grenoble 1 / CNRS-INSU, Institut de Plan\'etologie et d'Astrophysique de Grenoble (IPAG) UMR 5274, Grenoble, F-38041, France\\
$^{4}$ Aix Marseille Universit\'e, CNRS, LAM (Laboratoire d'Astrophysique de Marseille) UMR 7326, 13388, Marseille, France\\
$^{5}$ Institut d'Astrophysique de Paris, UMR7095 CNRS, Universit\'e Pierre \& Marie Curie, 98bis boulevard Arago, 75014 Paris, France\\
$^{6}$ Observatoire de Haute-Provence, Universit\'e d'Aix-Marseille \& CNRS, 04870 Saint Michel l'Observatoire, France\\
$^{7}$ CNRS, Canada-France-Hawaii Telescope Corporation, 65-1238 Mamalahoa Hwy., Kamuela, HI 96743, USA\\
$^{8}$ Departamento de F\'isica e Astronomia, Faculdade de Ci\^encias, Universidade do Porto, Portugal\\
}
\begin{document}

\date{Accepted 2015 May 11.  Received 2015 May 11; in original form 2015 April 1}

\pagerange{\pageref{firstpage}--\pageref{lastpage}} \pubyear{2015}

\maketitle

\label{firstpage}

\begin{abstract}
The statistical validation of transiting exoplanets proved to be an efficient technique to secure the nature of small exoplanet signals which cannot be established by purely spectroscopic means. However, the spectroscopic diagnoses are providing us with useful constraints on the presence of blended stellar contaminants. In this paper, we present how a contaminating star affects the measurements of the various spectroscopic diagnoses as function of the parameters of the target and contaminating stars using the model implemented into the \texttt{PASTIS} planet-validation software. We find particular cases for which a blend might produce a large radial velocity signal but no bisector variation. It might also produce a bisector variation anti-correlated with the radial velocity one, as in the case of stellar spots. In those cases, the full width half maximum variation provides complementary constraints. These results can be used to constrain blend scenarios for transiting planet candidates or radial velocity planets. We review all the spectroscopic diagnoses reported in the literature so far, especially the ones to monitor the line asymmetry. We estimate their uncertainty and compare their sensitivity to blends. Based on that, we recommend the use of BiGauss which is the most sensitive diagnosis to monitor line-profile asymmetry. In this paper, we also investigate the sensitivity of the radial velocities to constrain blend scenarios and develop a formalism to estimate the level of dilution of a blended signal. Finally, we apply our blend model to re-analyse the spectroscopic diagnoses of HD16702, an unresolved face-on binary which exhibits bisector variations.
\end{abstract}


\section{Introduction}

One of the main objectives of detecting new extrasolar planets is to constrain theories of planet formation, migration and evolution \citep[e.g.][]{2009A&A...501.1161M}. Exoplanet discoveries are therefore used to build up our general knowledge about planetary systems. It is therefore crucial that these exoplanets are indeed \textit{bona-fide} planets. However, periodic signals with a relatively low amplitude compatible with a planetary nature are not necessarily planets \citep{2012Natur.492...48C}. Indeed, many astrophysical non-planetary scenarios (the so-called ``false-positives'') can mimic the signal of a planet. All techniques are affected: the radial-velocity planetary signals might be mimicked by stellar activity \citep[e.g.][]{2001A&A...379..279Q} or nearly face-on binaries \citep{2012A&A...538A.113D, 2013ApJ...770..119W} ; the photometric transit signals of a planet can be mimicked by various configurations of eclipsing binaries \citep[e.g.][]{2011ApJ...738..170M, 2013A&A...557A.139S} ; false-positive scenarios might also produce misinterpreted direct imaging and microlensing exoplanet detections \citep[][respectively]{2013A&A...554A..21Z, 2013ApJ...778...55H}. In some cases, the source of the signal is a planet, but the amplitude of the signal is diluted by the presence of a brighter star in the system \citep[e.g.][]{2002A&A...392..215S}. Therefore, the derived properties of the planets (without accounting for the contaminating light) might be significantly different, hence limiting the accuracy of statistical analysis based on planet detections \citep[e.g][]{2012ApJS..201...15H, 2013PNAS..11019273P, 2013ApJ...766...81F}.\\

It is therefore fundamental to secure the planetary nature of a detection. This is especially important for transiting exoplanets, for which a substantial fraction of transit-like events are not of planetary origin \citep[e.g.][]{2009A&A...506..337A, 2009A&A...506..321M, 2012A&A...545A..76S, 2014AJ....147..119C}. One way to establish the planetary nature of a transiting candidate is through high-resolution spectroscopy and radial velocity measurements \citep{2003ApJ...597.1076K}. Spectroscopic observations can efficiently screen out eclipsing binaries as well as triple systems \citep{2012A&A...545A..76S}. Then, to establish the planetary nature of a candidate, it is commonly considered that a transit candidate has to satisfy all the following requirements: (1) a radial velocity variation is significantly detected in phase with the transit ephemeris ; (2) the inferred mass is compatible with that of a planet ; (3) no significant correlation is found between this RV variation and the asymmetry of the line profile (the bisector) ; (4) the signal is stable in time and amplitude \citep{2014A&A...566A..35S}. \\

This technique works relatively well for systems where the reflex motion of the host can be observed with current instrumentation. However, for the thousands of potential small planets that the \textit{Kepler} space telescope detected \citep[e.g.][]{2013ApJS..204...24B}, the reflex motion is too small to be observed \citep{2013sf2a.conf..555S}, therefore limiting the establishment of their planetary nature. Other non-spectroscopic methods exist to measure the mass of a transiting planet, such as the relativistic Doppler-boosting effect \citep{2010A&A...521L..59M} and through dynamical interaction with other planets in the system \citep{2010Sci...330...51H}. However, these techniques are limited to very short orbital period planets or multiple-planet systems \citep[respectively]{2011AJ....142..195S, 2012MNRAS.422L..57B}. \\

An alternative solution has been proposed by \citet{2011ApJ...727...24T}, called the planet-validation. This technique consists in evaluating the probability of each scenario to produce the observed data. If the planet scenario turns out to be significantly the most likely scenario, then the planet is considered as validated statistically. However, as demonstrated by \citet{2014MNRAS.441..983D}, if the transit signal has a signal-to-noise ratio below 150, it is not possible to validate the nature of a transiting exoplanet based only on the photometric data. In such configuration, the validation relies on the accuracy of the \textit{a priori} information. This is clearly a limitation for the exploration of planets like the Earth since the statistics in this population is extremely low. To overcome this limitation, the solution is to increase the statistical weight of the data compared with the priors in the validation procedure, by modelling data from different techniques providing different constraints to the various scenarios.\\

The objective of this paper is to show how spectroscopic diagnoses can be used to constraints false-positive scenarios in the context of the validation of planets performed with the \texttt{PASTIS} software\footnote{Planet Analysis and Small Transit Investigation Software} \citep{2014MNRAS.441..983D}. For that, we explore the effects of a blended star on the measurement of both the radial velocity, the full width half maximum (FWHM) and the line-asymmetry diagnoses already defined in the literature (e.g. the Bisector Inverse Slope). These diagnoses are commonly used to disentangle planetary signals from stellar activity in radial velocity data \citep[see e.g.][and reference therein]{2012Natur.491..207D}. These diagnoses have already been used by, e.g. \citet{2002A&A...392..215S}, \citet{2004ApJ...614..979T}, \citet{2005ApJ...619..558T}, \citet{2012A&A...538A.113D}, \citet{2013ApJ...770..119W}, and recently by \citet{2014A&A...571A..37S} to analyse blended radial velocity of double-line binary systems mimicking planets or brown dwarfs. \\

Note that the impact of stellar contamination on precise radial velocity has already been studied by \citet{2013A&A...550A..75C} in the context of the preparation of the ESPRESSO spectrograph. Previously, the effect of a contaminating star has also been studied by \citet{2005ApJ...619..558T} to constrain the false-positive scenario of OGLE-TR-56 which mimic both a transit and radial velocity signals of a planet. In this paper, we extend the work done by these authors, including all the spectroscopic diagnoses. We also investigate how these diagnoses change as function of the parameters of the target and contaminating star. Finally, we propose a formalism to estimate the amplitude of a radial velocity signal as function of the parameters of the blended components.\\

We first explore in Section \ref{ContImpact} the impact of a blended contaminant on precise radial velocity and the line-profile diagnoses. Then, we present in Section~\ref{sensitivity} the sensitivity of radial velocity data to constrain blend scenarios. In Section~\ref{RealCase}, we test our \texttt{PASTIS} RV-blend model by re-analysing an unresolved nearly face-on binary, HD16702. We report some caveats in Section~\ref{Caveats}. Finally, we draw our conclusions in Section~\ref{Discut} and discuss how these radial velocity constraints complement the photometric ones in the context of planet validation.\\

Several appendices are available at the end of this paper. In Appendix \ref{RVdiag}, we review how to compute the various spectroscopic diagnoses used in this paper. In Appendix \ref{CCFModel}, we describe our method to simulate the line profile of a blended stellar system. In Appendix \ref{ComputTime}, we present the computation time as well as the numerical precision and accuracy to measure each line-profile diagnosis. Finally, in Appendix \ref{Anal16702}, we provide the parameters used to analyse the system HD16702 with \texttt{PASTIS}, together with their prior and posterior distributions.

\section{Effect of a contaminating star on the spectroscopic diagnoses}
\label{ContImpact}

\subsection{Presentation of the spectroscopic diagnoses}
\label{sectDiagnPres}
At the resolution of most spectrographs used to measure precise radial velocities of stars (40 000 -- 120 000), an unblended and unsaturated weak line profile of a slow-rotating star could be approximated by a Gaussian function. Averaging thousand of such lines using the cross-correlation technique \citep[hereafter CCF][]{1996A&AS..119..373B, 2002A&A...388..632P} also produces a Gaussian profile. The CCF presents the advantage of having a much higher signal-to-noise ratio compared with individual lines without modifying the shape of the averaged line profile. This function is characterised by a given location in the radial velocity space, a contrast and a FWHM. An asymmetry in the observed line profile can be produced by, e.g. the convective blue shift, pulsations, stellar spots \citep{2014ApJ...796..132D}, and contaminating stars. To monitor the line asymmetry, several diagnoses have been defined in the literature. They are detailed in the Appendix \ref{RVdiag} of this paper and illustrated in Fig. \ref{DiagnFig}. For clarity, we list them here:

\begin{itemize}
\item BIS \citep{2001A&A...379..279Q}: it consists in computing the velocity span between the average of the top portion of the bisector and the bottom portion of the bisector, within some defined limits.
\item \vspan\, \citep{2011A&A...528A...4B}: it consists in fitting a Gaussian profile to the top and bottom part of the line profile. The asymmetry diagnosis is defined as the difference between the two velocities.
\item BiGauss \citep{2006A&A...453..309N}: it consists in fitting an asymmetric Gaussian profile to the line profile (see eq. \ref{AsymGauss}). The asymmetry diagnosis is defined as the difference between the radial velocity found with the symmetric and asymmetric Gaussian profiles.
\item \vasy\, \citep{2013A&A...557A..93F}: it consists in computing the difference of spectral information \citep[as defined by][]{2001A&A...374..733B} contained in the blue and red wings of the line profile.
\item \wspan\, (this work, see Appendix \ref{RVdiag}): it consists in fitting a Gaussian profile to the red and blue wings of the line profile. The asymmetry diagnosis is defined as the difference between the two velocities.
\end{itemize}

These diagnoses can be measured directly on each individual stellar lines available in the spectrum, as long as their profiles have been correctly reproduced by the spectrograph and are not perturbed by an external source like other unresolved lines, or the spectrum of a calibration lamp, an iodine cell, the Earth atmosphere, or the Sun light reflected by the Moon. To improve their precision, those diagnoses can be measured on the averaged line profile which is obtained by computing the weighted cross-correlation function \citep{1996A&AS..119..373B, 2002A&A...388..632P}. Note that this technique conserves the line-profile shape only if the observed spectrum is correlated with a binary, box-shaped mask \citep{2010EAS....41...27E}. The correlation with a non-box-shaped mask such as a master-spectrum template \citep{2012ApJS..200...15A} or a synthetic model might smooth out the diagnoses variation, if any, and thus decrease the efficiency of the method presented here.

\subsection{Blend simulation}
\label{blendsimu}

To test the effect of a contaminating star, we simulated, as described in Appendix \ref{CCFModel}, the line profile of a target star with a radial velocity of 0 \kms, a FWHM of 10\kms, and a contrast of 30\%. Those values are typical for a main-sequence star with solar metallicity and a slow rotation observed at high-resolution (R $\gtrsim$ 40000). We then blended this main line profile with a contaminating one with a FWHM of 70\%, 100\%, and 130\% of the target FWHM, a radial velocity shift, $\phi$, from the target star ranging from $\phi=0$\kms\ to $\phi=35$\kms\ and a contrast of 30\%. We fixed a flux ratio between the target and the blend lines of 1 to 100. Those values have been chosen to illustrate the effect of a blend on precise spectroscopic diagnosis. Changing their value will not change the general behavior presented here, it will just rescale all the values. For each pair of radial velocity shift and FWHM of the blended star, we measured the radial velocity, contrast, FWHM as well as the five line-asymmetry diagnoses presented in Section \ref{sectDiagnPres} and in Appendix \ref{RVdiag}. The results are displayed in Fig. \ref{AllDiagn}. The dash, solid and dot-dash lines represent the measured diagnoses as function of the radial velocity shift between the two stars for a FWHM of the blend of 7\kms, 10\kms, and 13\kms, respectively.\\

\begin{figure*}
\centering
\includegraphics[width = \textwidth]{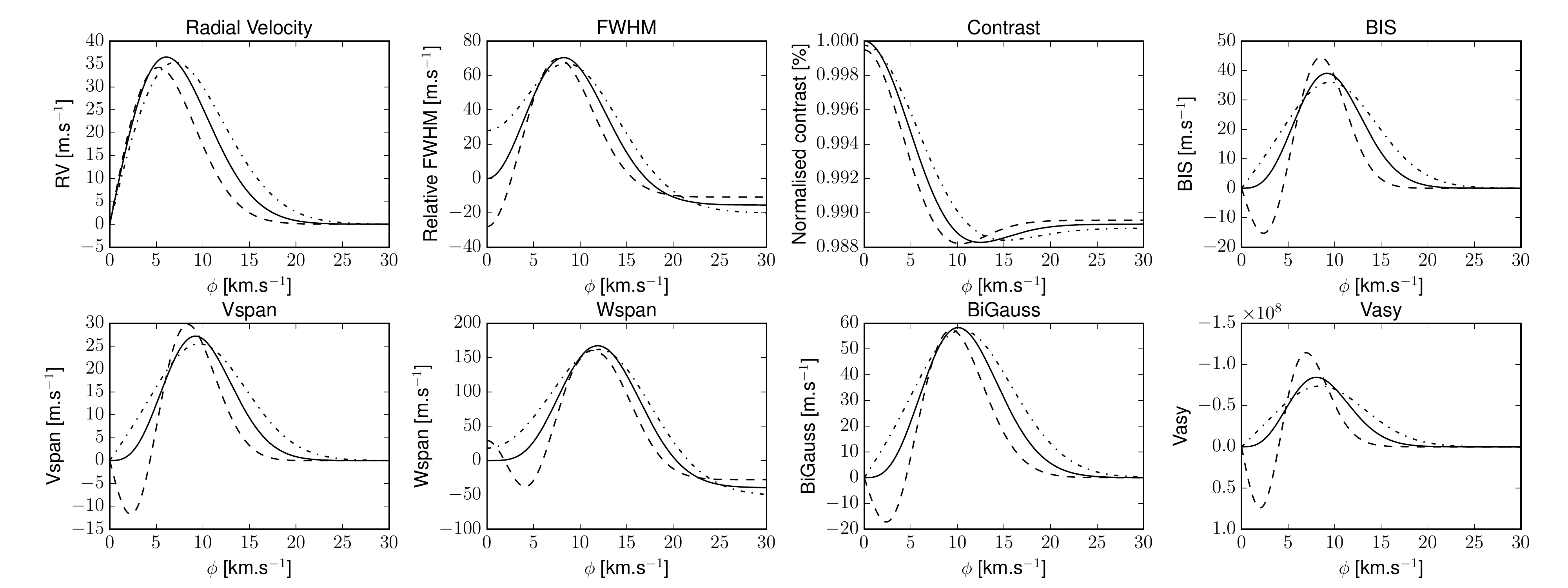}
\includegraphics[width = \textwidth]{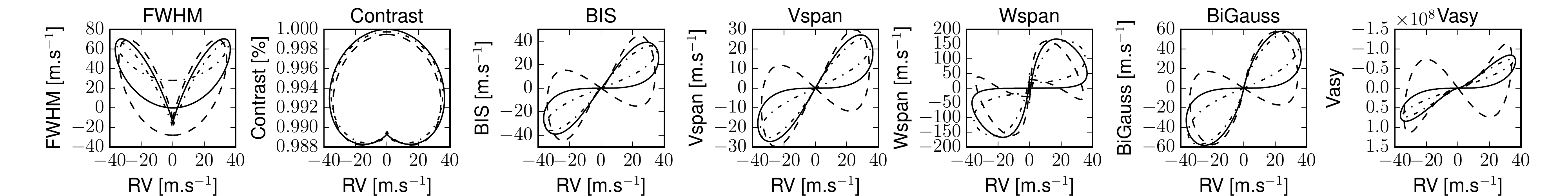}
\caption{Top and middle plots: Measured RV, FWHM, contrast, BIS, Vspan, Wspan, BiGauss, Vasy (from left to right and top to bottom) as function of the RV offset between the target and the contaminating stars ($\phi$, see text), for a flux ratio in the continuum between the two stars of 1\%. Bottom plots: Measured correlation between the radial velocity and FWHM, contrast, BIS, Vspan, Wspan, BiGauss, Vasy (from left to right). The dash, solid and dot-dash lines represent the measured diagnoses as function of the radial velocity shift between the two stars for a FWHM of the blend of 7\kms, 10\kms, and 13\kms\ (respectively).}
\label{AllDiagn}%
\end{figure*}

We also check the correlation between the measured diagnoses and the measured radial velocity. The correlation plots are displayed in Fig. \ref{AllDiagn}. We discuss the results of these simulations in the Sections \ref{EffectRV} -- \ref{EffectBIS}.

\subsection{Effect on the radial velocity}
\label{EffectRV}
The effect of a contaminating star on the measurement of the radial velocity is displayed in Fig. \ref{AllDiagn}. Such perturbation is qualitatively similar to the Rossiter-McLaughlin anomaly, where a faint fraction of the line profile is ``subtracted'' by the eclipsing object. This Rossiter-McLaughlin anomaly has been described analytically, e.g. in \citet[eq. 15]{2013A&A...550A..53B}, assuming Gaussian profiles and a small eclipsing object (i.e. a planet) compared with the stellar radius. On the opposite, here, the faint line of a contaminant is added to the target one. If we assume that both the target and the contaminant have Gaussian profiles and a flux-ratio small enough (i.e. the contaminant is much fainter than the target), we can approximate the RV-blend anomaly behavior as the opposite of the Rossiter-McLaughlin anomaly. Thus, following \citet{2013A&A...550A..53B}, we can describe the radial velocity-blend anomaly ($\xi_{RV}$) by the equation:
\begin{equation}
\label{RVperturb}
\xi_{RV}\left(\phi\right) = \left(\frac{2\sigma_{1}^{2}}{\sigma_{1}^{2}+\sigma_{2}^{2}}\right)^{3/2} \phi f \exp\left\{-\frac{\phi^{2}}{2\left(\sigma_{1}^{2}+\sigma_{2}^{2}\right)} \right\},
\end{equation}
where $\sigma_{1}$ and $\sigma_{2}$ are the Gaussian widths of the two line profiles, which depends on the \vsini and the instrumental resolution, $\phi$ is the radial velocity shift between them and $f$ is the flux ratio. \\

Using equation \ref{RVperturb}, we can determine the radial velocity shift that maximize the perturbation ($\phi_{max}$) by computing the zero of its first derivative:
\begin{equation}
\label{RVmax}
\frac{\partial\xi_{RV}\left(\phi\right)}{\partial\phi} = 0 \Longrightarrow \phi_{max} = \pm \sqrt{\sigma_{1}^{2}+\sigma_{2}^{2}}.
\end{equation}

Therefore, a contaminating star affects the measurement of the radial velocity in two different ways, depending if $|\phi|$ is lower or greater than $|\phi_{max}|$. If $|\phi| < |\phi_{max}|$, the observed radial velocity will be \textit{correlated} with the radial velocity variation of the blending star. On the opposite, if $|\phi| > |\phi_{max}|$, the observed radial velocity will be \textit{anti-correlated} with the radial velocity variation of the contaminating star. Note that by definition of the Rayleigh criterion, if $|\phi| > |\phi_{max}|$, the two line profiles are spectroscopically \textit{resolved} and if $|\phi| < |\phi_{max}|$, they are spectroscopically \textit{unresolved}.  \\

Now, if we compute the maximum amplitude of $\xi_{RV}$ when $\phi = |\phi_{max}|$ in equation \ref{RVperturb}, we find that:
\begin{equation}
\label{AmpRVanomaly}
\xi_{RV}\left(\phi_{max}\right) = 2\sqrt{2}\frac{\sigma_{1}^{3}}{\sigma_{1}^{2}+\sigma_{2}^{2}}f\exp\left\{-\frac{1}{2}\right\}.
\end{equation}
This equation shows that the larger $\sigma_{1}$ is, due to stellar rotation or to lack of instrumental resolution, the larger is the RV-blend anomaly signal.\\

The RV-blend anomaly displayed in Fig. \ref{AllDiagn} is similar to the one found by \citet{2013A&A...550A..75C}, except that we do not found the secondary anomaly reported by these authors at large RV-separation, which is caused when the contaminant leaves the cross-correlation domain. For larger separation, the lines are resolved and this technique does not provide constraints to planet validation. 

\subsection{Effect on the FWHM}
\label{effectFWHM}
The FWHM is affected by the presence of a contaminant. In the cases that we simulated here, the variation in FWHM is about two times larger than the one in radial velocity (see Fig \ref{AllDiagn}). As for the radial velocity, the FWHM varies in case of blended star in two different regimes: the FWHM is correlated with $\phi$ for small RV separation and anti-correlated for large RV separation. The maximum of the FWHM-blend anomaly occurs at a different value of $\phi$. If the two stars have similar radial velocities ($\phi \approx 0$), a small radial velocity variation of the contaminant will not produce a large variation of the FWHM value. The variation is even fainter if the contaminant star is rotating fast. 

\subsection{Effect on the line contrast}
\label{EffectContrast}
The contrast of the line profile is also affected by the contaminating stars. Compared with the other diagnoses, the contrast presents only small variations. Moreover, the contrast can be perturbed by several instrumental systematics such as residuals of the CCD dark current correction, non-linearity of the CCD, sky luminosity variation caused by the Moon, twilight, or city lights or even sky transparency related to the humidity and the presence of clouds. Therefore, the contrast of the line profile is not a good diagnosis to constrain blended components. The contrast might still be used to highlight unseen stars in a spectrum (e.g. O-type stars) which would not present many stellar lines apart from the Balmer lines, but would dilute substantially the spectrum of the contaminant. In this case, a narrow line profile with a low contrast could be observed, and the low-contrast anomaly might be used to reveal the presence of a massive or very fast-rotating star. However, in extremis, this could also be mimicked by a metal-poor star. For these reasons, we will not mentioned it anymore in the rest of the paper.

\subsection{Effect on the line asymmetry}
\label{EffectBIS}
It is well known that a contaminating star affects the asymmetry of the line of the target star if their radial velocity separation is small enough. All the asymmetry diagnoses present a similar anomaly curve, with different amplitude, except for \vasy\, which present an anomaly curve with an opposite sign (note the reversed Y axis in Fig. \ref{AllDiagn} for the \vasy\, diagnosis). This opposite shape of the \vasy\, anomaly curve is due to its different definition compared with other diagnoses (see appendix \ref{RVdiag} for more details about their definition). Compared to the other diagnoses, we observe three different regimes of the asymmetry-anomaly curves presented in Fig. \ref{AllDiagn}, depending on the FWHM of the contaminating star (FWHM$_{2}$), compared with the target star (FWHM$_{1}$):
\begin{enumerate}
\item \textbf{FWHM$_{1} <\ $ FWHM$_{2}$:} if the contaminating star is rotating faster than the target star (see the dot-dash line in Fig. \ref{AllDiagn}), the anomaly curves of the BIS, \vspan\, \wspan, BiGauss and \vasy\, present a shape that is similar with the radial velocity-anomaly curve. The main difference is that the maximum anomaly is reached for a radial velocity separation of about twice $\phi_{max}$ (see Fig. \ref{AllDiagn}). Such separation difference in the maximum of the anomalies is the cause of the ``figure eight loop'' seen in the correlation between the line asymmetry and the radial velocity (see Fig. \ref{AllDiagn}). This shape of the BIS -- RV correlation plot has already been observed by \citet{2011A&A...528A...4B} for stellar activity, but with an opposite sign (see for example their figure 6). \\

\item \textbf{FWHM$_{1} \approx\ $ FWHM$_{2}$:} If the two stars have similar FWHMs (see the solid line in Fig. \ref{AllDiagn}), the asymmetry diagnoses are not very sensitives to a contaminant with a small radial velocity separation. This configuration could be quite common when the FWHM of both stars are dominated by the instrumental resolution and not by stellar rotation. A contaminating star with the same FWHM than the target star and with a radial velocity variation up to a few \kms\ around the target systemic radial velocity will only slightly affect the BIS, \vspan, \wspan, BiGauss, and \vasy. In this blind zone, the observed radial velocity and asymmetry diagnoses might \textit{not} produce a significant correlation (see also Section \ref{Diluted}).\\ 

\item \textbf{FWHM$_{1} >\ $ FWHM$_{2}$:} If the target star is rotating faster than the contaminating star (see the dash line in Fig. \ref{AllDiagn}), all asymmetry diagnoses present three regimes: for small values of $\phi$, the measured asymmetry is anti-correlated with $\phi$ and thus, with the observed radial velocity. Then, for intermediate values of $\phi$, the asymmetry diagnoses are correlated with the observed radial velocity. Finally, as in case (i), when the two stars have a large RV separation, the asymmetry diagnosis is anti-correlated with $\phi$ but correlated with the observed radial velocity.
\end{enumerate}

\subsection{Photon-noise uncertainty}
\label{photnoise}

The various asymmetry diagnoses presented in the previous section are not of equal sensitivity to blend scenarios. The diagnosis that presents a maximum of sensitivity to a contaminant is \wspan, then Bigauss, BIS, and finally \vspan. This has to be compared with the sensitivity to noise in the line profile. For that, we estimated the photon noise uncertainty in the various diagnoses. We took a solar spectrum observed on the blue sky with a signal to noise (SNR) greater than 150 by the HARPS \citep{2003Msngr.114...20M} and SOPHIE \citep{2008SPIE.7014E..17P, 2009A&A...505..853B} spectrographs. For SOPHIE, we used two different spectra, one for each instrumental configuration: the high-resolution (HR, R $\sim$ 75000) and high-efficiency (HE, R $\sim$ 39000) modes. For comparison, HARPS has a resolution of R $\sim$ 110000. We add white noise to these spectra at the level of 10\%, 5\%, 2\%, and 1\% of the flux in each pixel, that we generated 100 times. We ended with 400 spectra for each instrument. We cross correlated all these spectra using a G2V numerical mask and measured all the diagnoses. We then computed the standard deviation of all the diagnoses for each noise level. They all exhibit a clear correlation with the radial velocity photon noise. We fitted this correlation with a 1-D polynomial and report in the Table \ref{ErrorDiags} the value of the slope, for each diagnosis and each instruments. Therefore, these values should be used to scale the diagnoses uncertainty, given the radial velocity photon noise such as:
\begin{equation}
\sigma_k = \varepsilon_{k} \sigma_{RV},
\end{equation}
with $\sigma_k$ the photon noise uncertainty of the diagnosis $k$, $\varepsilon_{k}$ the scaling coefficient listed in Table \ref{ErrorDiags}, and $\sigma_{RV}$ the radial velocity uncertainty. Note however that those values represent only the photon noise uncertainty. Those diagnoses are also sensitives to instrumental effects \citep[e.g.][]{2010A&A...523A..88B, 2011EPJWC..1602003B, 2012A&A...538A.113D, 2014A&A...571A..37S} which might produce important additional red noise. Note that the typical photon noise reached by SOPHIE and HARPS on stars in the magnitude range 12 -- 16 is presented in \citet{2011EPJWC..1102001S} and compared with the expected radial velocity amplitude of the \textit{Kepler} candidates in \citet{2013sf2a.conf..555S}.\\

\begin{table}
  \centering 
  \caption{Photon noise uncertainty on the spectroscopic diagnoses, relative to the one of the radial velocity, expressed in \kms. The contrast is expressed here in \%.}
\setlength{\tabcolsep}{0.9mm}
  \begin{tabular}{lccccccc}
\hline
Spectrograph & FWHM & BiGauss & BIS & \vspan & \wspan & \vasy & Contrast\\ 
\hline
     HARPS  &  2.0  &  2.1  &  2.0  &  1.5  &  5.8  &  5.6 10$^{10}$  &  11.3 \\
 SOPHIE HR  &  2.5  &  2.3  &  2.3  &  1.6  &  6.8  &  3.6 10$^{10}$  &  10.2 \\
 SOPHIE HE  &  2.5  &  2.7  &  2.6  &  1.8  &  8.4  &  2.5 10$^{10}$  &  7.0 \\
\hline
\end{tabular}
  \label{ErrorDiags}
\end{table}

We find that the photon noise uncertainty of the FWHM is about a factor of two bigger than the RV, as for BIS and BiGauss. This agrees with the values usually assumed \citep[e.g.]{2009A&A...506..303Q}. As claimed by \citet{2011A&A...528A...4B}, \vspan\, is less noisy than BIS by a factor of about 25\%. However, the new line-asymmetry diagnosis we present in this paper is less robust to noise than BIS, its photon noise uncertainty is larger by a factor of about 3. The dimension-less diagnosis \vasy\, present a high-value uncertainty, which is difficult to interpret. By comparing the amplitude of the \vasy-blend effect in figure \ref{AllDiagn} with the one of the radial velocity, we realised it exhibits a variation $\Delta$\vasy\, / $\Delta$RV $\approx$ 3.10$^{9}$ km$^{-1}$.s larger than the one in radial velocity. However, its photon noise uncertainty scaling coefficient, $\varepsilon_{\rm V_{\rm asy}}$, is at the level of a few 10$^{10}$ km$^{-1}$.s. Therefore, the diagnosis \vasy\, seems to be more sensitive to the noise than to line-profile asymmetry. Based on that, we do not recommend to use \vasy\ to monitor line-profile asymmetries. This is apparently not similar for the contrast, which exhibits in the simulation of Section \ref{blendsimu} a RV-normalised variation at the level of $\Delta$contrast / $\Delta$RV =  34\%.km$^{-1}$.s while its photon noise scaling coefficient is at the level of $\varepsilon_{\rm contrast}$ = 10\%.km$^{-1}$.s.  However, the contrast is affected by numerous systematic effects discussed in Section \ref{EffectContrast}, which make this diagnosis not reliable for the purpose of this paper.

\subsection{Comparison of the diagnoses sensitivity}

\begin{figure}
\begin{center}
\includegraphics[width = \columnwidth]{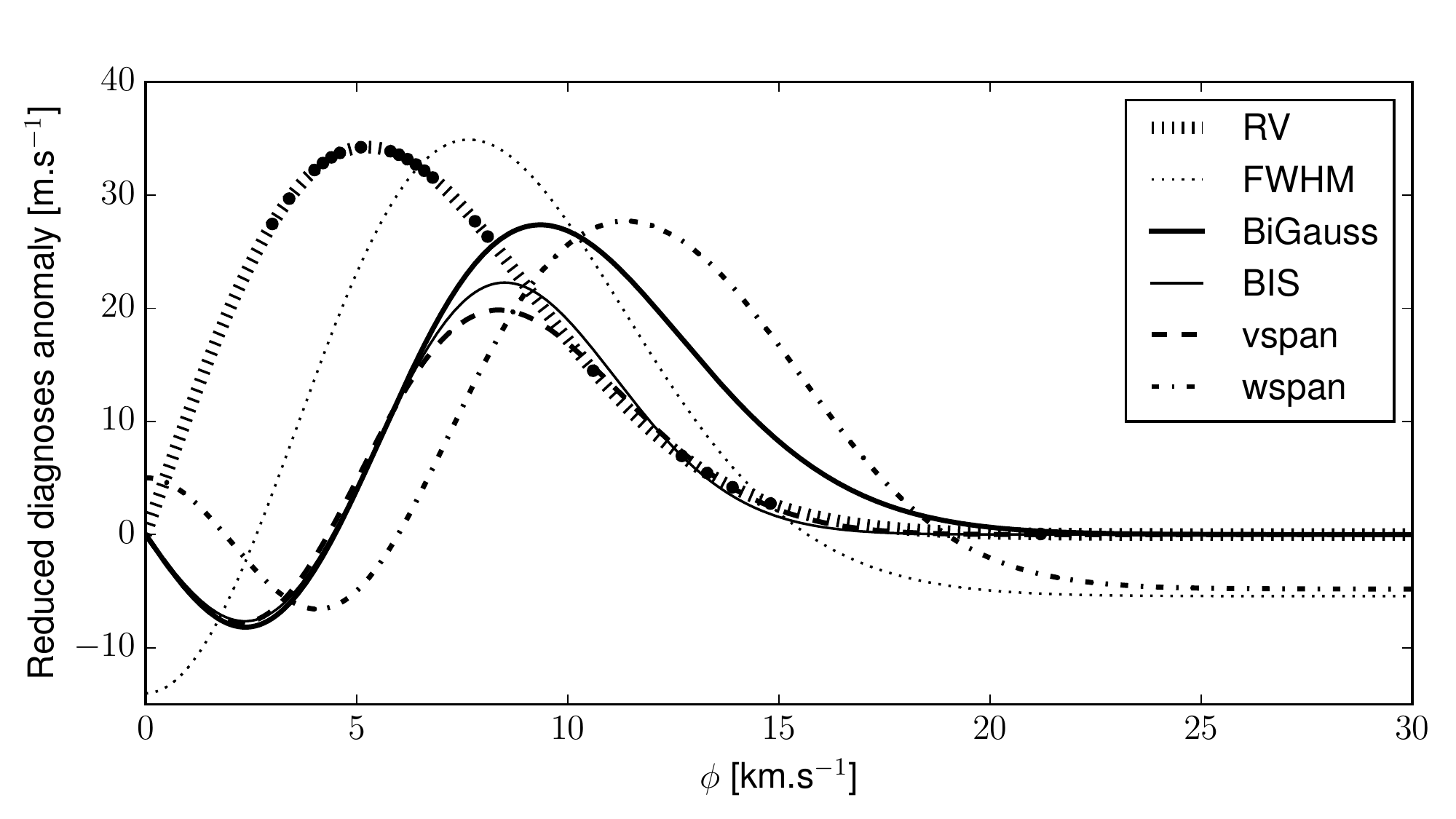}
\caption{Comparison of the blend curves for all diagnoses but \vasy\, presented in this paper. The simulation is the one described in Section \ref{blendsimu}, for a contaminating star with a FWHM of 7\kms. The amplitude of the anomaly curves has been divided by the photon-noise coefficients for HARPS from Table \ref{ErrorDiags}.}
\label{DiagnComp}
\end{center}
\end{figure}

We report five different line-asymmetry diagnoses that present different sensitivity to blend and to the photon noise. To compare them, we normalised the anomaly curves produced in Section \ref{blendsimu} with the photon-noise coefficients of the Table \ref{ErrorDiags}. We find that the most sensitive diagnoses to line-profile asymmetry are BiGauss and \wspan\, (see Fig. \ref{DiagnComp}). They are followed by BIS and \vspan. Being a dimension-less diagnosis, \vasy\, is difficult to compare with the other ones. However, as discussed in the previous section, \vasy\, is more sensitive to noise than line-profile asymmetry. Thus, we do not consider it further in the end of the paper. Compared to \wspan\, and \vspan, BiGauss does not need a high oversampling of the line profile for their fitting procedure to be stable (see Appendix \ref{ComputTime}). BiGauss is also the fastest one to compute at a given fit precision. Even if we did not compared those diagnoses in the case of stellar activity, we expect the same results concerning their sensitivity to line-profile asymmetry. For all these reasons, we recommend to use BiGauss to monitor line-profile asymmetry. \\

We also see in Fig. \ref{DiagnComp} that the RV-, FWHM-, and asymmetry diagnoses exhibit different blend anomaly curves, except when the two stars are completely separated spectroscopically. The FWHM and the line-profile asymmetry diagnoses do not have a zero slope for the same value of $\phi$. Therefore, by combining both the FWHM and the line-profile asymmetry diagnoses, one should be able to identify unambiguously a blended contaminant, if enough precision on those diagnoses is reached.

\subsection{Illustration with a planet orbiting a companion star}
\label{Diluted}

To illustrate the effect of a contaminating star, we arbitrary simulated a binary system composed by a primary star of 1\Msun\, and a secondary star of 0.6\Msun\, in a long-period orbit. In the V band, the flux ratio between the two stars is of about 1:30 for main-sequence stars with solar metallicity. We assumed a systemic RV shift between the two stars of $\phi_{0}=1$ \kms. We then assumed that the secondary star has a dark companion that produces a RV variation with a semi-amplitude $K_{2} = 50 \ms$ (this corresponds to a Jupiter-mass planet at 200 days or to a Saturn-mass planet at 5 days). As in the previous section, we simulated the observation of such system assuming that the primary star has a FWHM$_{1}$ of 10 \kms\ and the secondary has a FWHM$_{2}$ of 7 \kms\ (dash line), 10\kms\ (solid line), and 13\kms\ (dot-dash line). For simplicity in this example, we considered a circular orbit. The blended radial velocities of this system present a nearly circular apparent signal with a semi-amplitude of about 1.5\ms. The dilution of the signal is discussed in Section \ref{RVdilution}.\\

\begin{figure}
\begin{center}
\includegraphics[width = \columnwidth]{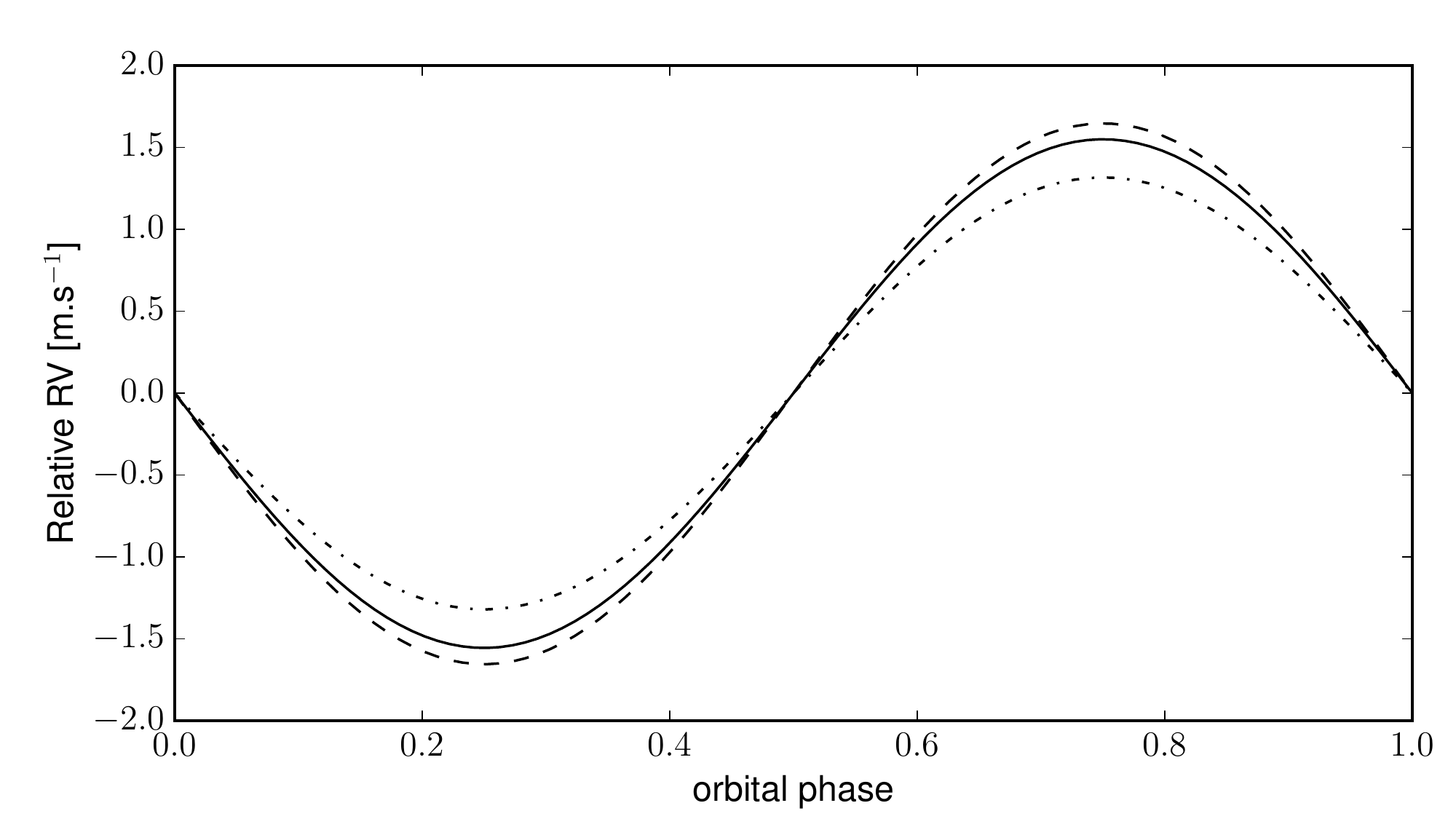}
\includegraphics[width = \columnwidth]{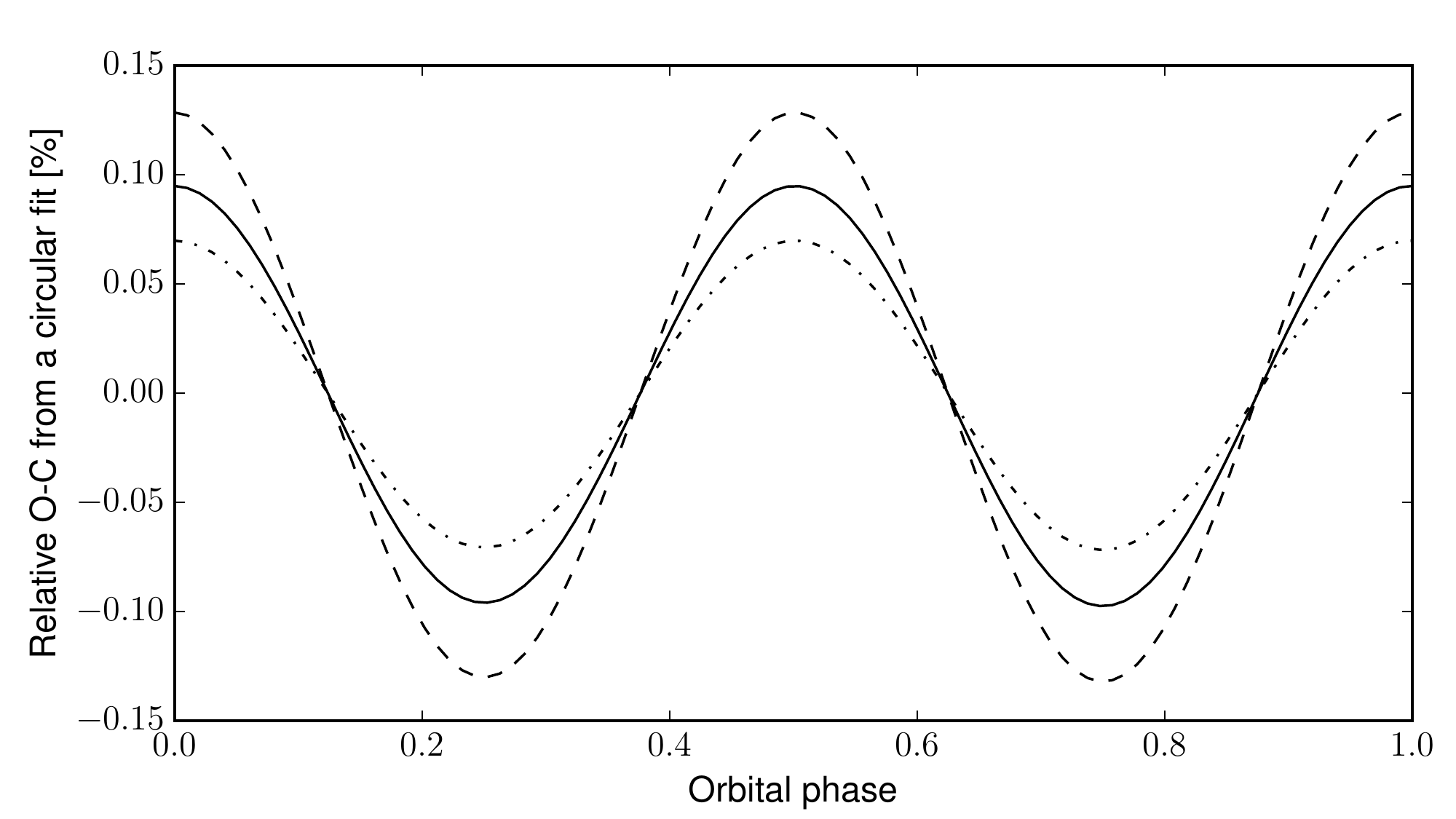}\\
\includegraphics[width = \columnwidth]{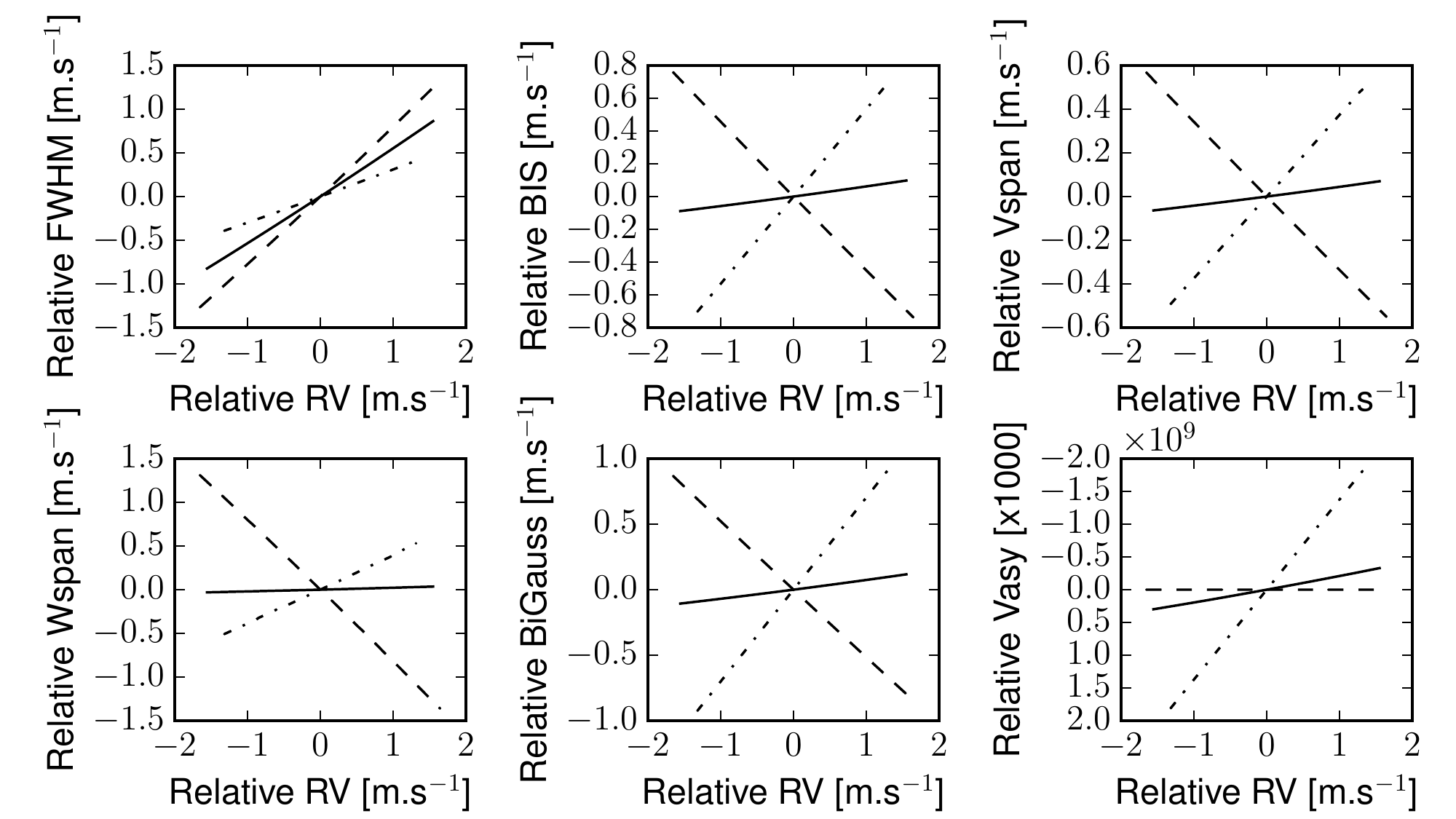}\\
\caption{Simulation of a planet orbiting a companion star (see text). \textit{Upper panel}: blended radial velocities as function of the orbital phase and relative to their systemic velocity. \textit{Middle panel}: relative residuals of the blended radial velocities from a circular fit as function of the orbital phase of the companion transiting planet. The residuals exhibit a signal at half the initial period. \textit{Lower panels}: correlation between the measured RV and the FWHM, BIS, \vspan, \wspan, BiGauss and \vasy\ (from middle to bottom and left to right). The different lines represent different FWHM$_{2}$: 7\kms\ (dash line) 10\kms\ (solid line), and 13\kms\ (dot-dash line).}
\label{exampleBlend1}
\end{center}
\end{figure}

The blended signal is nearly but not strictly circular, as already pointed out by \citet{2013ApJ...770..119W}. As shown in the middle panel of Fig. \ref{exampleBlend1}, the residuals from a circular fit exhibit another signal exactly at half of the initial period. In the present simulation, this residual signal has an amplitude of about 0.1\% of the one of the blended signal (hence 0.15 cm.s$^{-1}$). This P/2 residual signal should be maximum for $\phi_{0}=0$ and $K_{2} = \phi_{max}$ (eq. \ref{RVmax} -- here $K_{2} \approx$ 6 \kms\, which would correspond to a 5\Mjup brown dwarf at 5 days). In this particular case, the blended radial velocity signal exhibits an amplitude of about 100\ms\, and the residual signal has a relative amplitude of 6\% (hence 6\ms, for FWHM$_{1,2} = 10$ \kms). If $K_{2} > \phi_{max}$ or if the reflex motion of the secondary star does not always satisfy $|\phi| < \phi_{max}$ along the orbit, then the blended signal is clearly not a Keplerian, as already illustrated in \citet{2005ApJ...619..558T}. Note that, as found by \citet{2013ApJ...770..119W}, this residual signal could also be at the second harmonic, hence at P/3.\\

We also display in Fig. \ref{exampleBlend1} (bottom panels) the correlations between the relative RV and the relative FWHM, BIS, \vspan, \wspan, BiGauss and \vasy. We do not display here the contrast since it presents only a variation at the level of $\sim$ 0.3\%, which is undetectable with current instrumentation. The blended signal revealed a positive-slope correlation between RV and BIS, \vspan, \wspan\ and BiGauss for FWHM$_{1} < $ FWHM$_{2}$ and a negative-slope correlation otherwise. If the two stars present a similar FWHM, then the correlation between the aforementioned line-asymmetry diagnoses and the RV is very weak, which might be undistinguishable from noise. Only the FWHM might present a clear correlation with the RV in this case. This shows that it is possible to have a false positive scenario which produce the radial velocity variation compatible with a planet but no correlation between the RV and the line-profile asymmetry diagnoses. It is also possible that a false-positive scenario produce a negative-slope correlation between these two quantities, which might be misinterpreted as stellar activity. In both cases, the FWHM correlates with the RV. Thus, the FWHM is a good diagnosis that should be used in complement to other line-asymmetry diagnoses to constrain blend scenarios and infer the planetary nature of a signal.

\section{Sensitivity of radial velocities to blends}
\label{sensitivity}

\subsection{Dilution of the radial velocity signal}
\label{RVdilution}

\begin{figure}
\begin{center}
\includegraphics[width = \columnwidth]{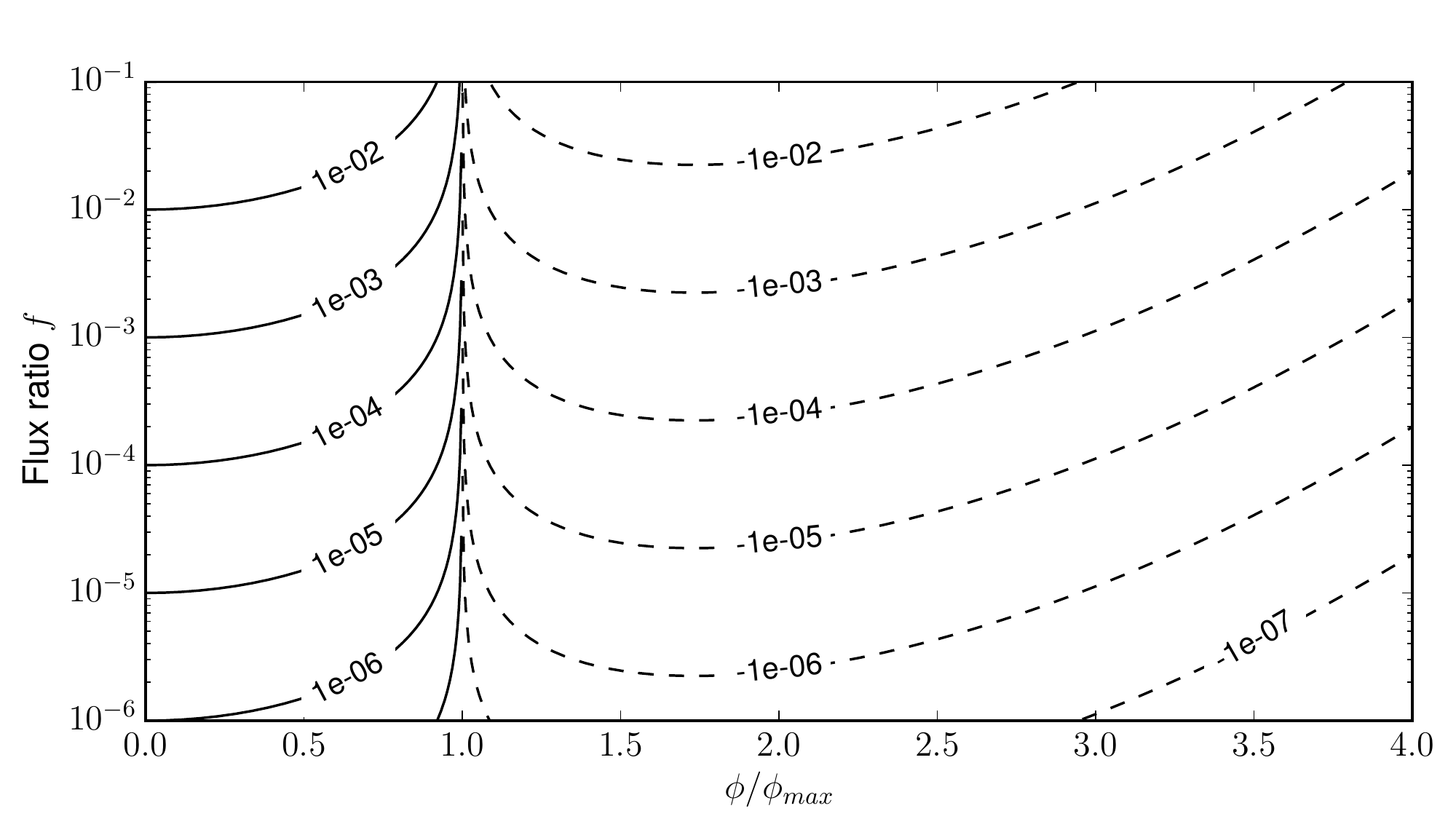}
\caption{Iso-contours of the RV dilution $\partial\xi_{RV} / \partial\phi$ as function of the normalised RV shift between the main and secondary stars, $\phi/\phi_{max}$ and the flux ratio between their line profile, $f$. Solid lines represent positives values of $\partial\xi_{RV} / \partial\phi$ while dash lines stand for negative values. Flux ratios greater than 0.1 are not displayed here because they do not respect the approximation used in this computation (see text). For this figure, we assumed $FWHM = \sigma_{1}\times2\sqrt{2\ln(2)} = 10$\kms.}
\label{figdilution}
\end{center}
\end{figure}

The presence of a blended contaminant dilutes any radial velocity variation \citep{2011ApJS..197....3B, 2013A&A...550A..75C}. As exemple, we simulated in the previous section a 50\ms-signal which was seen, after dilution by another star, as being less than 2\ms. To quantify this dilution in radial velocity, we evaluated the ratio between the amplitude of the blended RV variation (i.e. the one that would be observed) and the real one. In the approximation of low flux ratio and Gaussian profiles, we can use eq. \ref{RVperturb} to estimate the blended RV variation, $\Delta\xi_{RV}$, given the RV variation of the contaminating star, $\Delta\phi$ (i.e. a variation of RV offset). In other words, $\Delta\xi_{RV}$ is the radial velocity variation that would be observed while $\Delta\phi$ is the true one. We can therefore describe the radial velocity dilution factor $\Delta\xi_{RV} / \Delta\phi$ as:

\begin{eqnarray}
\frac{\Delta\xi_{RV}}{\Delta\phi} &\approx& \frac{\partial\xi_{RV}}{\partial\phi}\; \mathrm{for}\; \Delta\phi \to 0 \label{dilution}\\
\frac{\partial\xi_{RV}}{\partial\phi} &=& 2\sqrt{2}\left(\frac{\sigma_{1}}{\phi_{max}}\right)^{3}f\left(1-\left(\frac{\phi}{\phi_{max}}\right)^{2}\right)\exp\left(-\frac{1}{2}\left(\frac{\phi}{\phi_{max}}\right)^{2}\right) \label{dilution2}
\end{eqnarray}
The figure \ref{figdilution} displays the iso-contours of $\partial\xi_{RV} / \partial\phi$ as function of $\phi/\phi_{max}$ and $f$, for $FWHM = \sigma_{1}\times2\sqrt{2\ln(2)} = 10$\kms. Note that these analytical results are perfectly reproduced by our numerical simulation for all $f \ll 1$. Three regimes are clearly visible:
\begin{itemize}
\item $\phi \ll \phi_{max}$: if the two line profiles are unresolved, the RV dilution scales as $\Delta\xi_{RV} / \Delta\phi \propto f$.\\
\item $\phi \approx \phi_{max}$: if the two line profiles are shifted exactly by $\phi_{max}$, a small variation of the contaminating star would \textit{not} result in a RV variation of the blended line profile. In this case, the radial velocity is \textit{insensitive} to faint blend variations.\\
\item $\phi > \phi_{max}$: if the two line profiles are resolved, the RV dilution scales as $\Delta\xi_{RV} / \Delta\phi \propto - f$. This means that the blended RV variation will be \textit{anti-correlated} with the contaminant variation. If a particular epoch of the orbit is constrained by other means (e.g. the transit epoch is known thanks to photometric data), it would result in a blended RV variation in anti-phase \citep[as reported among the \textit{CoRoT} detections, e.g.][]{2009A&A...506..501C, 2012A&A...539A..14E, 2012A&A...538A.112C, 2012Ap&SS.337..511C}, or with a negative amplitude \citep[e.g.][]{2014ApJS..210...20M}.
\end{itemize}

We assume now the opposite case, where the brightest component has a RV variation $\Delta\phi^{\prime}$ and the faintest component has a constant RV in time. In the referential of the brightest star, this RV variation will be seen as a RV variation of the secondary star $\Delta\phi$ such as $\Delta\phi^{\prime} = - \Delta\phi$, hence the RV dilution will be:

\begin{equation}
\frac{\Delta\xi_{RV}^{\prime}}{\Delta\phi^{\prime}} = 1 - \frac{\Delta\xi_{RV}}{\Delta\phi} \approx 1 - \frac{\partial\xi_{RV}}{\partial\phi}\; \mathrm{for}\; \Delta\phi \to 0,
\end{equation}
with $\Delta\xi_{RV}^{\prime}$ the blended radial velocity variation given a real variation of the main star of $\Delta\phi^{\prime}$. The same approximations are used in this equation than for eq. \ref{dilution}. We validated this analytical expression with a numerical simulation. Therefore, following the three regimes displayed in figure \ref{figdilution}, we find that if a planet host is contaminated by a secondary star, hence the blended RV variation will be:
\begin{itemize}
\item smaller than the real variation if the two stars are unresolved ($\phi < \phi_{max}$) ;\\
\item unaffected if the two stars are shifted exactly by $\phi_{max}$ ;\\
\item greater than the real variation if the two stars are resolved ($\phi > \phi_{max}$).\\
\end{itemize}

An illustration of the first case is the WASP-85~A \& B binary system which hosts a transiting planet \citep{2014arXiv1412.7761B}. In this system, the unblended HARPS radial velocities present a \textit{larger} amplitude than the blended SOPHIE and CORALIE data. Note however that the formalism developed here can not be applied to this case since the flux ratio between the two stars is not low enough. This system could have been resolved though numerical simulation, as done in the \texttt{PASTIS} software, but the expected variation of BiGauss and FWHM is at the level of 1.5\ms\, and 10\ms\, for SOPHIE (respectively), which is much below the precision on this star. Other illustrations of radial velocity dilution by an unresolved contaminant is Kepler-14 \citep{2011ApJS..197....3B} and KOI-1257 \citep{2014A&A...571A..37S}. As discussed in the later paper, since nearly half the stars are in binary system, there might be much more diluted radial velocity cases than we have found so far.

\subsection{Radial velocity sensitivity to physical companions}

In the previous section, we presented the dilution of a given radial velocity signal as function of the flux ratio and the RV shift between the blended stars. To interpret this radial velocity dilution as a sensitivity to blends, we related the flux ratio to the physical properties of the blended stars. In that case, we tested the sensitivity of radial velocities to a companion star, i.e. bounded with the primary star. For this scenario, we considered a primary star with a mass of M$_{1} = 1$\Msun\, and a secondary one with M$_{2}\in [0.1;1.0]$ \Msun. We set that both stars have solar properties (\met\, = 0 dex, age = 5Gyr), the same distance, a \vsini of 3\kms, and a systemic radial velocity shift of $\phi_{0} = 4$\kms. We simulated a faint radial velocity variation of the secondary star, caused by the presence of a sub-stellar, non-emitting, companion. The RV variation ($\Delta\phi$) of this secondary star is kept relatively small (we assumed a circular 1-\Mjup\, companion at 365 days of period, hence K$_{2} < 130$\ms). \\

We tested this scenario for the SOPHIE spectrograph \citep{2008SPIE.7014E..17P, 2009A&A...505..853B}, for which the line profile has been calibrated by \citet{2010A&A...523A..88B}. We used these calibrations (also reported in Appendix \ref{CCFModel}) to simulate this system for the two instrumental configurations: the High-Efficiency mode (HE) with a spectral resolution of $\sim 39000$ and the High-Resolution mode (HR) with a spectral resolution of $\sim 75000$. We also considered the two masks G2 and K5 that are used to compute the cross-correlation function in the online pipeline. For each M$_{2}$, instrumental mode and mask, we simulated the CCFs of both stars, add them, and measure the blended RV variation $\Delta\xi_{RV}$ that we divided by $\Delta\phi$. For that, we used the Dartmouth evolutionary tracks \citep{2008ApJS..178...89D} to estimate the stellar parameters (radius, \teff, luminosity) and the BT-SETTL \citep{2012IAUS..282..235A} spectral atmosphere models to estimate the flux of both stars within the spectrograph bandpass (assumed to be similar to Johnson-V) and their (B-V) color index needed to derive their CCF parameters (see Appendix \ref{CCFModel}). \\

We display in figure \ref{ConstraintsRV} (upper panel) the ratio $\Delta\xi_{RV} / \Delta\phi$ as function of the mass of the bounded secondary star, host of the sub-stellar companion, for both instrumental modes and correlation masks. In this configuration and with a 1\ms\, precision in the radial velocity we can constrain the signature of a blended K$_{2} = 100$\ms\, RV signal on a companion with a mass down to $\sim$ 0.5\Msun. To detect the same signal on a 0.3\Msun, one needs a precision at the level of a few 10\cms.
As expected, the blended RV variation decreases as the secondary star become smaller and thus fainter in the V-band (the flux ratio decreases). We observe that for M$_{2} \ll $ M$_{1}$, the radial velocity signal is larger in the mask K5 than in the mask G2. This is the so-called mask effect already reported in \citet{2009IAUS..253..129B}. The K5 mask has more lines than the G2 one. Thus, the CCF of a blended G2+K5 spectrum will present a slightly larger flux ratio if correlated using all the lines of a K5 star compared with the lines of a G2 star. Since a larger flux ratio implies a larger blended signal, the K5 mask is more sensitive to K and M dwarfs than the G2 mask.  The difference of radial velocity as derived by the two masks therefore provide constraints on the mass of the contaminant star compared with the one of the primary star.\\

In the figure \ref{ConstraintsRV}, we also observe that the HE mode of SOPHIE is more sensitive to blends than the HR mode. In eq. \ref{dilution2}, the factor $\partial\xi_{RV}/\partial\phi$ scales as $1-(\phi/\phi_{max})^{2}$ for $\phi < \phi_{max}$ (unresolved components). Since the instrumental resolution of the HE mode is lower than the HR one, the Gaussian widths of the stars are larger in HE than in HR, which makes $\phi_{max}$ larger, and thus $\partial\xi_{RV}/\partial\phi$ is also larger. Therefore, decreasing the instrumental resolution increases asymptotically the sensitivity of radial velocities to unresolved stellar components. On the other hand, for $\phi \gg \phi_{max}$ (i.e. resolved stellar components), the factor $\partial\xi_{RV}/\partial\phi$ is dominated by the term $\exp(-1/2(\phi/\phi_{max})^{2})$. In that case, increasing the instrumental resolution also increases the sensitivity of radial velocities to resolved stellar components.\\

To constrain the presence of a stellar contaminant in radial velocity data, a good test is either to use spectrographs with different instrumental resolution or to artificially decrease the instrumental resolution by convolving the spectra with a Gaussian profile of increasing width. Then, if the amplitude of the radial velocity signal varies significantly with the instrumental resolution, the line-profile is thus blended. Note that the opposite is not always true: since this is an asymptotic effect, no amplitude variation as a function of the instrumental resolution does not imply that the line profile is unblended. At the most extreme case, when the two stellar components have perfectly the same radial velocity ($\phi = 0$), the factor $\partial\xi_{RV}/\partial\phi$ scales as $(\sigma_{1}^{2}/(\sigma_{1}^{2} + \sigma_{2}^{2}))^{3/2}$. So, in that case, if both $\sigma_{1}$ and $\sigma_{2}$ are dominated by the instrumental resolution and not by the stellar rotation or if both stars have the same \vsini, the factor $\partial\xi_{RV}/\partial\phi$ is independent of the instrumental resolution. However, if only one star is rotating fast compared with the instrumental resolution, this test should be conclusive, even if the stars have a similar systemic radial velocity.\\

Note that high-precision spectrographs in the near infrared, like SPIRou \citep{2013sf2a.conf..497D} or CARMENES \citep{2014SPIE.9147E..1FQ}, should also provide further constrain on the presence of low-mass stars, by improving the flux ratio between the primary and the lower-mass companion stars.\\

\begin{figure}
\begin{center}
\includegraphics[width = \columnwidth]{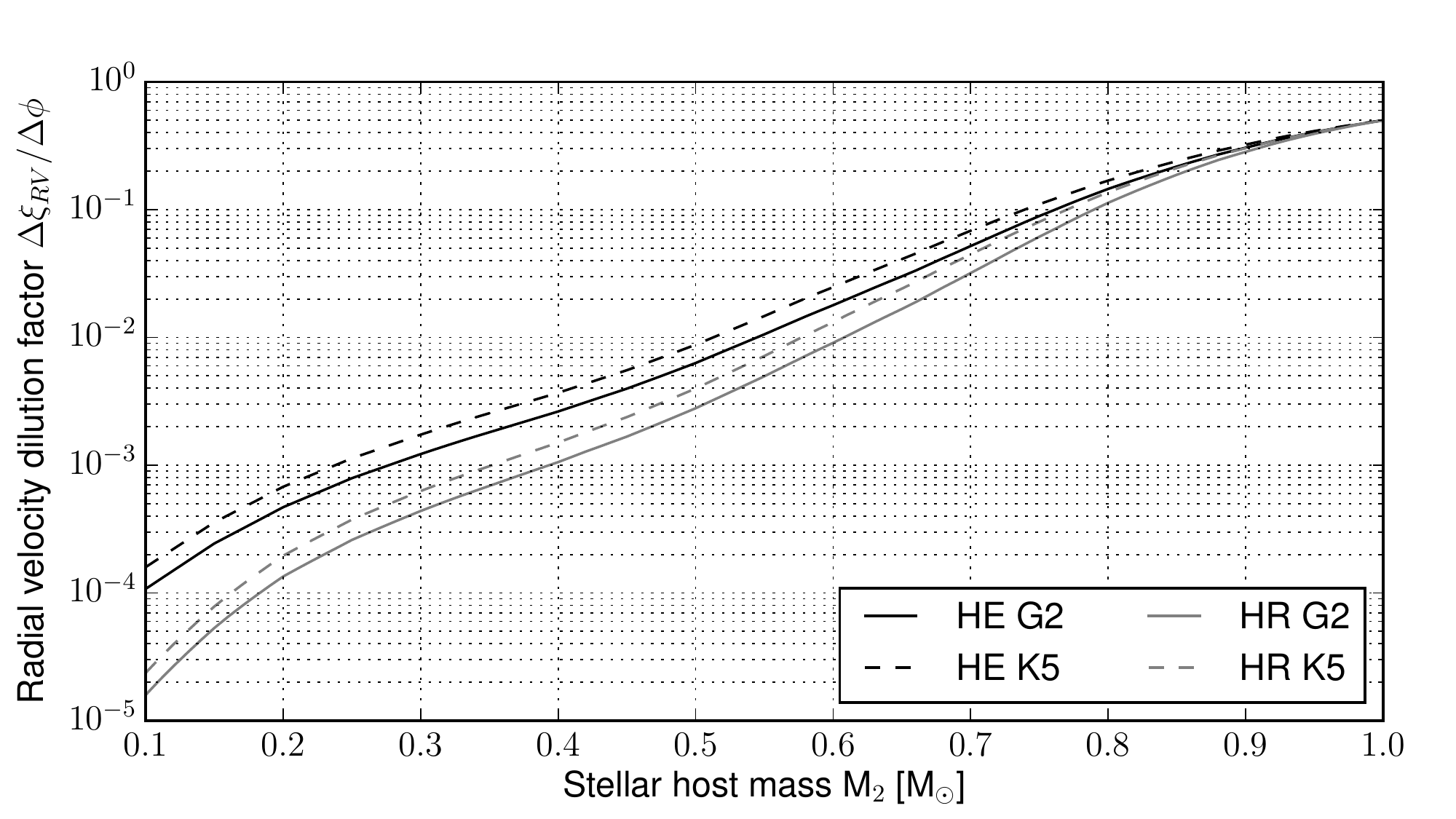}
\includegraphics[width = \columnwidth]{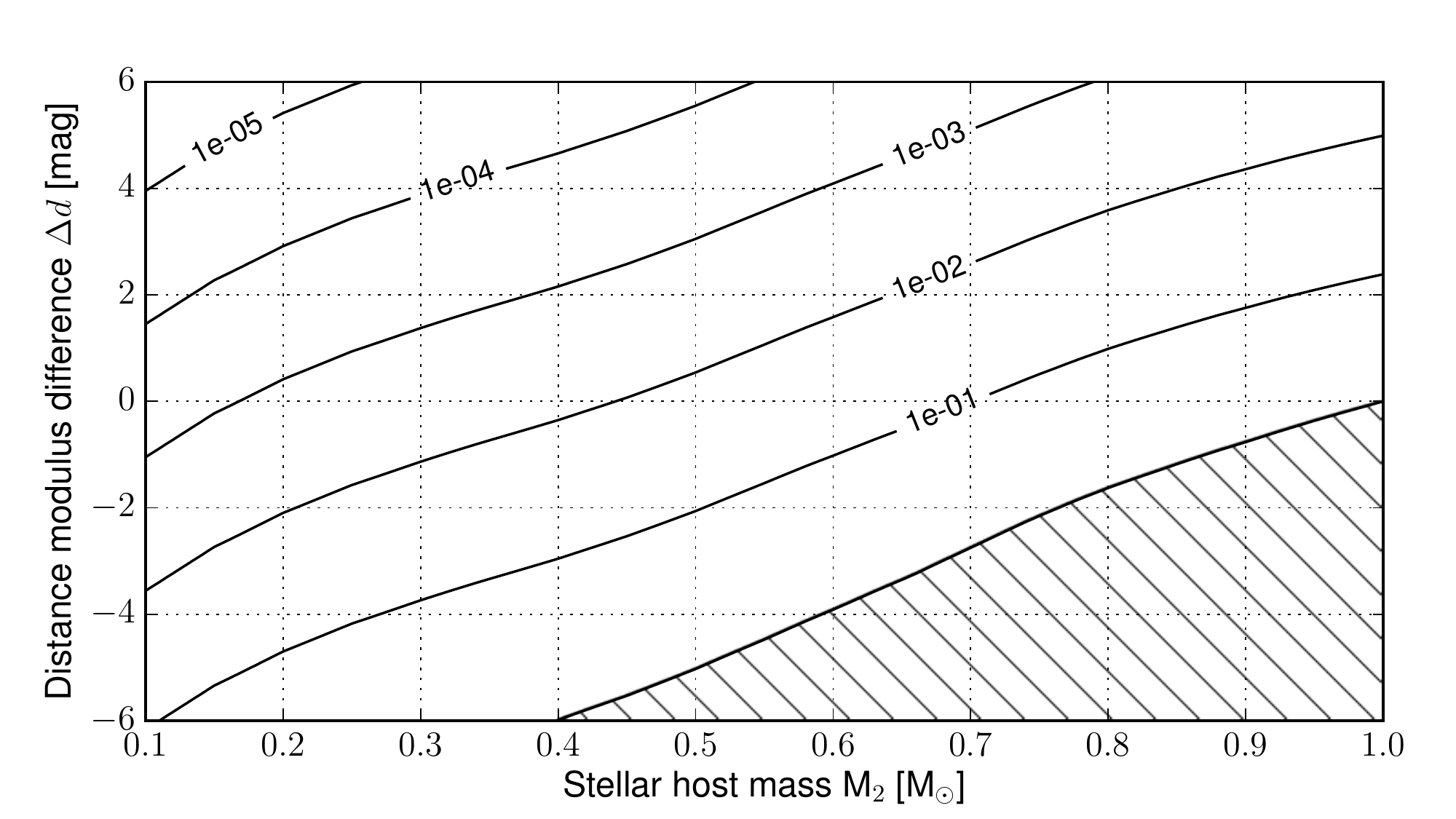}
\caption{Relative amplitude of a blended radial velocity variation, $\Delta\xi_{RV} / \Delta\phi$, as function of the mass of the blend host M$_{2}$ for the considered scenario (see text). \textit{Upper panel:} The four lines display this relative amplitude for both the SOPHIE HE (black lines) and HR modes (grey lines) using cross-correlation mask of a G2 (solid lines) and K5 dwarf (dash lines). This assumes $\phi_{0} = 4$\kms. \textit{Lower panel:} Iso-contours of this relative amplitude as function of the blend mass M$_{2}$ and the distance modulus difference $\Delta d$ (see text), for the SOPHIE HE mode and the G2 mask. The hatched region represents contaminant stars brighter than than the target in the V-band.}
\label{ConstraintsRV}
\end{center}
\end{figure}

\subsection{Radial velocity sensitivity to background companions}

A similar test can be done by relaxing the assumption that the primary and contaminating stars are bounded. This scenario has to assume that both stars are chance-aligned in both the plane of the sky and in the RV space, which is quite unlikely \citep{2013A&A...550A..75C}. Even though, we can quantify the constraints that the radial velocities could provide. For that we repeat the previous simulation by allowing the distance ratio between the two stars to vary in the range $\Delta d \in [-6 ; +6]$, with $\Delta d$ the distance modulus difference as defined in \citet{2011ApJS..197....5F}:
\begin{equation}
\label{DistLumi}
\Delta d = 5\log_{10}\left(\frac{D_{1}}{D_{2}} \right),
\end{equation}
where $D_{1}$ and $D_{2}$ are the distance of the primary and contaminating stars, respectively. For this test, we set $\phi_{0}$ to zero for simplicity, which gives optimistic values of the radial velocity constraints. The lower panel of figure \ref{ConstraintsRV} shows the iso-contours of $\Delta\xi_{RV} / \Delta\phi$ as function of the mass of the host star (i.e. the contaminating star) and the distance modulus difference $\Delta d$ for the G2 mask. Since $\phi_{0}=0$, both instrumental modes give the same constraints here. In this configuration, with a 1\ms\, precision on SOPHIE, one could constrain a perfectly chance-aligned system with K$_{2}=100$\ms\, on a 1\Msun\, host with $\Delta d \leq 5$ or a 0.7\Msun\, host with $\Delta d \leq 2$. This map of the radial velocity sensitivity to blends could be compared with the constraints on background systems provided, e.g. by transit photometry, if any. From the photometric constraints on the possible false positive parameters, one could estimate the expected amplitude of the radial velocity signal of such false positive using this technique. Then, it would be possible to optimise the radial velocity follow-up of transit candidates as function of the precision of the available spectrographs and their expected efficiency to further constrain, or reject, a given false-positive scenario. This could permit to minimize the needs of radial velocity spectrograph to screen out false positives of the future transit space missions \textit{TESS} \citep{2014SPIE.9143E..20R} and \textit{PLATO} \citep{2014ExA....38..249R}.\\

\section{Benchmarking the blend model: the case of HD16702, a nearly face-on binary}
\label{RealCase}

To validate our model of blended spectroscopic diagnosis, we tested it on one case which has other, independent constraints. This case is the system HD16702 revealed in \citet{2012A&A...538A.113D}. Observed by SOPHIE, this system exhibit a radial velocity signal compatible with a brown dwarf companion. A careful analysis of the line bisector only, with a preliminary version of this model, revealed that the system is a nearly face-on binary. The same authors also reported the marginal detection of an astrometric signature in the \textit{HIPPARCOS} data compatible with this scenario. Recently, \citet{2015AJ....149...18K} developed a new technique to detect faint components in the Keck/HIRES spectra. They applied this new technique to HD16702 and confirmed the bisector and astrometric results. In this section, we reanalyse the SOPHIE data of HD16702 with our blend model, by fitting simultaneously the radial velocities, FWHM, and BiGauss derived in the two correlation masks G2 and K5.\\

We assumed a scenario of a nearly face-on binary. We fitted all the data simultaneously using the \texttt{PASTIS} software \citep[see][and reference therein]{2014MNRAS.441..983D} which uses a Markov Chain Monte Carlo (MCMC) method. We used the Dartmouth evolutionary tracks \citep{2008ApJS..178...89D} to estimate the stellar parameters and the BT-SETTL models \citep{2012IAUS..282..235A} for the stellar atmospheric spectrum. The exhaustive list of parameters used for this analysis is reported in Table \ref{params16702}. We ran 15 chains randomly started from the joint prior distribution and all chains converged toward the same solution. We thinned and merged them to derive the posterior distribution. The derived value for the parameters and their 68.3\% uncertainty are also reported in Table \ref{params16702}. \\

\begin{figure}
\begin{center}
\includegraphics[width=\columnwidth]{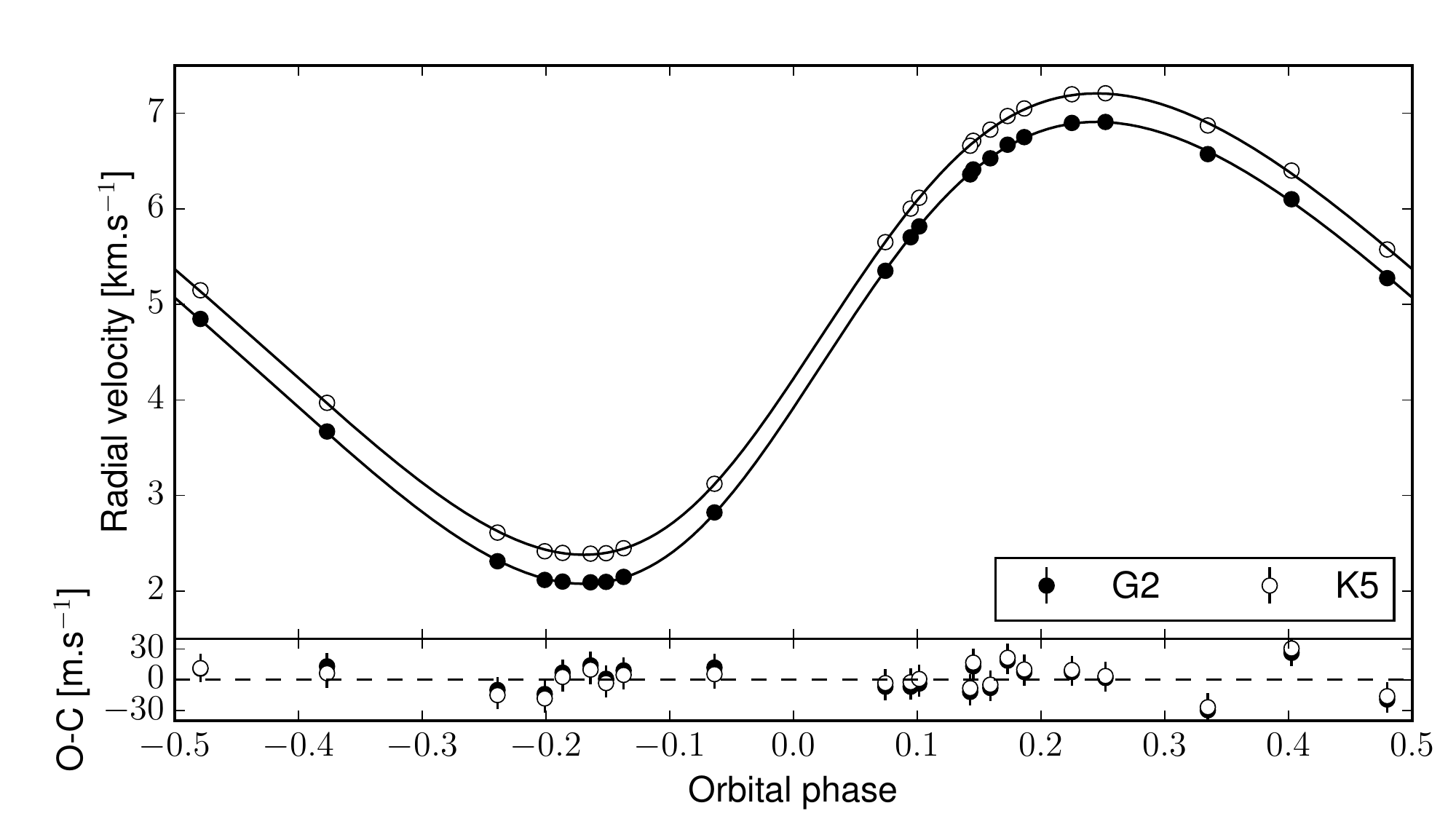}
\includegraphics[width=\columnwidth]{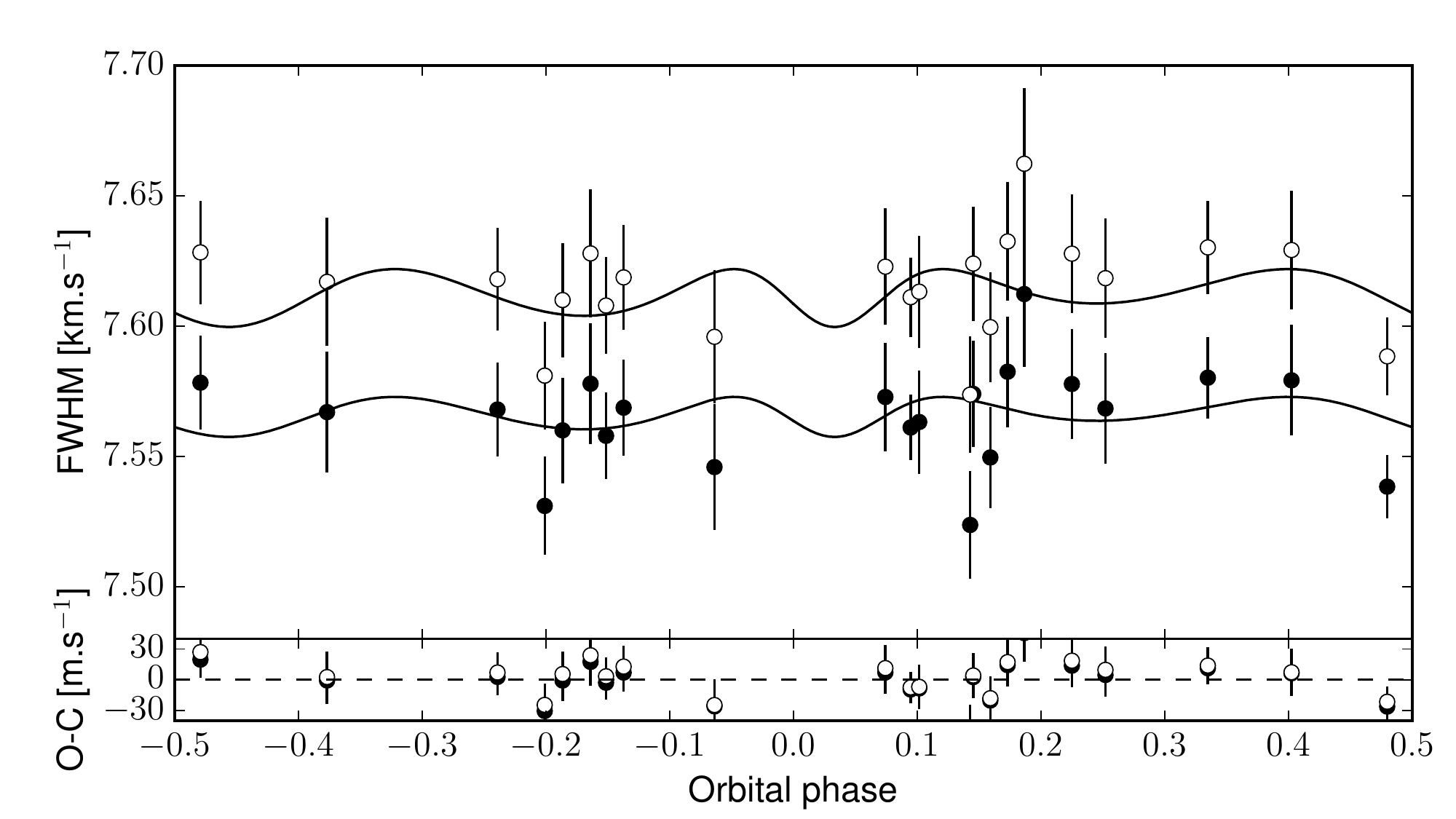} 
\includegraphics[width=\columnwidth]{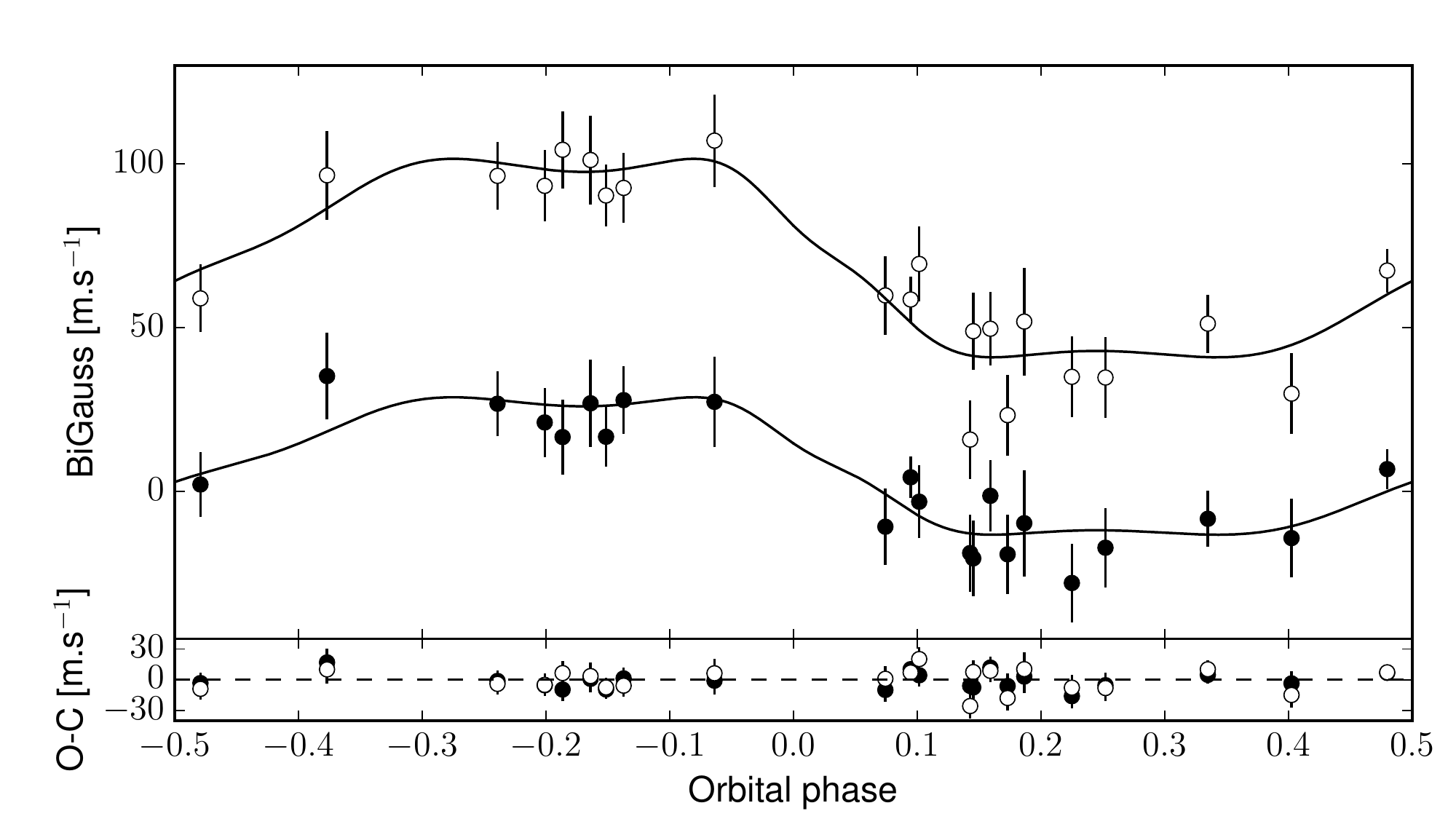}
\caption{Results from the analysis of SOPHIE data of the nearly face-on system HD16702. From top to bottom: phase-folded radial velocity velocities, FWHM, and BiGauss of this system, observed by the SOPHIE in the HR mode and derived using a G2 (black points) and K5 (open points) correlation mask. The data are plotted together with the best model found in the \texttt{PASTIS} analysis and their residuals. For clarity, an arbitrary offset has been set to the K5 data and models in the three plots.}
\label{16702}
\end{center}
\end{figure}

\begin{figure}
\begin{center}
\includegraphics[width=\columnwidth]{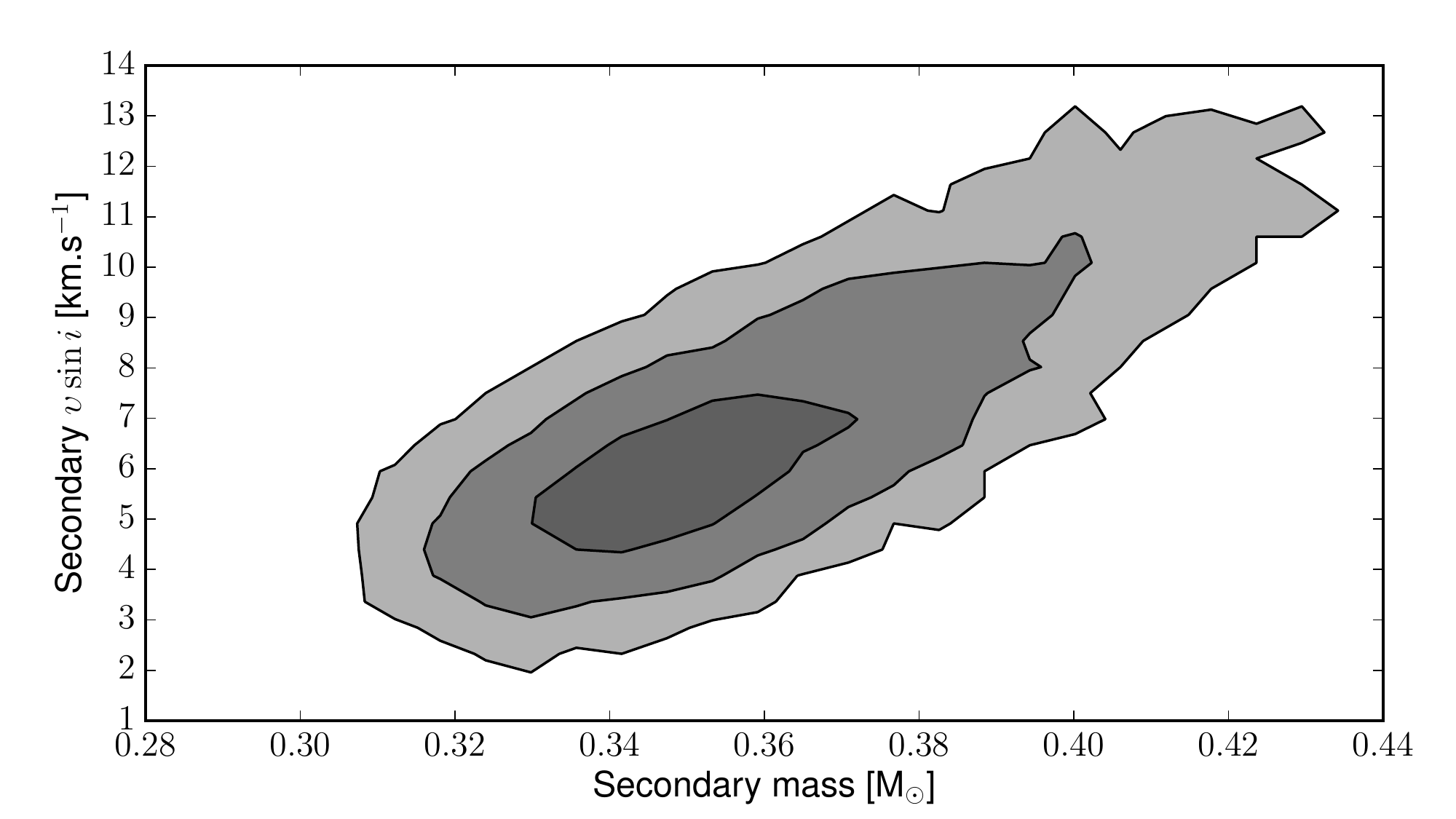}
\caption{Posterior distribution of the \vsini of the secondary star in the HD16702 system as function of its mass. The regions display the 68.3\%, 95.5\% and 99.7\% confidence intervals (from dark to light greys) on these parameters.}
\label{16702_bis}
\end{center}
\end{figure}

We display in figure \ref{16702} the phase-folded data considered in this analysis together with the best-fit model. While the radial velocity and BiGauss present a significant variation, the FWHM does not present a clear variation. This absence of significant variation is still used in the fitting procedure to constrain the \vsini of the secondary star. In \citet{2012A&A...538A.113D}, only the bisector was analysed and it was not possible to break the degeneracy between the FWHM of the secondary and its mass \citep[see the fig. 9 in][]{2012A&A...538A.113D}. In this analysis, our model was able to constrain the \vsini of the secondary star, as displayed in the figure \ref{16702_bis}. This was possible thanks to the additional constraints from the FWHM. We also ran the same analysis but for each correlation mask independently and found similar results, with slightly larger uncertainties as expected. Since the degeneracy between the mass of the secondary star and its \vsini is broken, we can further constraints the mass to a value of M$_{2} = 0.35 \pm 0.03$\Msun. Note that this uncertainty is only the statistical one. Uncertainties from the stellar models or from the blend model related to the calibration of the CCF are not taken into account here. This value is fully compatible with the values found by \citet{2012A&A...538A.113D} and \citet{2015AJ....149...18K} of M$_{2} = 0.40 \pm 0.05$\Msun\, and M$_{2} \sim 0.3$\Msun\, (respectively). This mass is also fully compatible with the one derived by the \textit{HIPPARCOS} astrometry of M$_{2} = 0.55 \pm 0.14$\Msun, which was used as prior for this analysis (see Table \ref{params16702}). Astrometric measurements from the \textit{GAIA} space telescope \citep{2012AN....333..453P} will permit us to confirm this result. This allows us to refine the inclination of the binary to $i_{orb} = 9.4 \pm 0.5$\degr. The RV-blend model we developed in the framework of the \texttt{PASTIS} validation tool is therefore able to reproduce perfectly the spectroscopic data of this blended, nearly face-on binary.

\section{Caveats}
\label{Caveats}
We would like to point out a few caveats in our model and results presented in this paper: 
\begin{itemize}
\item The calibration of the SOPHIE CCF that we are using is correct for (B-V) in the range of $[0.43 ; 0.98]$ while we are using it also for stars outside this domain. We assumed that it could be extrapolated without a significant impact on our results. We think this is not a limitation of our model because the parameters of the simulated line profile have only a second-order effect compared with the flux ratio. A new calibration of the SOPHIE CCFs with a wider range of (B-V) is mandatory to overcome this caveat. Note that calibrating the CCFs with masks corresponding to F and M dwarfs would improve the mask effect and thus the capability of this model to constrain higher or lower mass stars, respectively. Calibrating the CCF of other spectrographs in the visible but with different spectral resolution, like HARPS, HARPS-N, CORALIE, or in the near infrared could also improve the constraints. Note also that at a given (B-V) and \met, the surface gravity should also change the parameters of the CCF, which is not accounted for in the current calibration. The CCFs of evolved stars should also be considered for the calibration.\\
\item When we analyse the radial velocity, FWHM and line-asymmetry diagnosis from different correlation masks simultaneously within the \texttt{PASTIS} software, we compute the likelihood for each individual dataset and multiply them to get the marginal likelihood. This assumes that the dataset are independent from each other, which might not be the case. We assume however that this independence of the dataset, if any, does not affect significantly our results.\\
\item We model the line profile of the stars with a Gaussian profile. However, as soon as the stars are rotating faster than the instrumental resolution, their line profile differs significantly from a Gaussian profile. Our approximation might lead to over- or under-estimation of the spectroscopic diagnoses. Nevertheless, even if the values significantly change, we expect that the overall behaviour of the spectroscopic diagnoses is not affected by stellar rotation.
\end{itemize}

\section{Conclusion and discussion}
\label{Discut}

In this paper, we present the radial velocity model developed to constrain exoplanet blend scenarios with the \texttt{PASTIS} software. Using this model, we investigated how a blended stellar contaminant can perturb the measurement of the various spectroscopic diagnoses of a target star. We estimated the photon noise precision of these spectroscopic diagnoses which allowed us to conclude that BiGauss, as defined by \citet{2006A&A...453..309N}, is the most sensitive diagnosis to line-profile asymmetry.  By exploring the parameters of both the target and contaminating star, we find that a contaminant might produce different FWHM and bisector variations but similar radial velocity signals. In particular, if both stars have the same FWHM, their blended line profile does not exhibit strong asymmetry and can not be monitored with the bisector-like diagnoses. However, in that case, the FWHM should still present large variations that could be detected with stabilised high-resolution spectrographs. Therefore, the absence of bisector variation alone can not be used to firmly establish the nature of a planet detected by radial velocity. Both precise bisectors and FWHMs should not exhibit variation to exclude most astrophysical false-positive scenarios. \\

More interestingly, we find that if the FWHM of the contaminating star is smaller than the target one, the blended line bisectors will be anti-correlated with the radial velocities. This anti-correlation is usually produced by and interpreted as spots on the stellar surface \citep{2001A&A...379..279Q,2011A&A...528A...4B, 2012A&A...545A.109B, 2014ApJ...796..132D}. Therefore, a bona-fide planet in a blended binary could reproduce the signal in the RV, FWHM and bisector of an active star. Such system would produce a true-negative scenario, which might still be identified using other spectroscopic diagnoses like the variation of the $S$-index or of the $\log(\mathrm{R}^{\prime}_{\rm HK})$.\\

Our simulation also confirms the results of \citet{2013ApJ...770..119W}, that the blended signal of a circular radial velocity variation is not strictly circular. The residuals of a blended circular orbit exhibit another signal at the first or second harmonic (P/2 or P/3, respectively) with a relative amplitude up to a few percent of the initial blended signal. \\

In Section \ref{sensitivity}, we developed a new formalism to estimate the amplitude of the radial velocity dilution caused by a blended (brighter or fainter) star. It shows that, for example, a contaminating eclipsing binary spectroscopically unresolved with the target star (i.e. their RV shift is smaller than their joint profile widths) might mimic a radial velocity signal in phase with the eclipse ephemeris. On the other hand, if the two stars are resolved (i.e. their RV shift is greater than their joint profile widths), the blended radial velocity signal would be in anti-phase with the eclipse ephemeris. This formalism also revealed that if a radial velocity variation of a target star is blended with another fainter component, the later one would actually boost or dilute the amplitude of the signal, depending if the two stars are resolved or not (respectively). \\

Using this formalism, we also discussed the sensitivity of the SOPHIE spectrograph to detect physical and background companions to the target star. By comparing the two instrumental configurations of the SOPHIE spectrograph, we realised that the instrumental configuration with the lowest spectral resolution was more sensitive to unresolved contaminants than the configuration with the highest spectral resolution. This is however the opposite for a resolved contaminant. Therefore, by using different spectrographs with different spectral resolutions, one might constrain the presence of a contaminating star. Comparing the amplitude of signals from spectrographs with different spectral resolution provides another test to detect and constrain the presence of a stellar contaminant blended in the line profile of the target star. This can be done also by convolving the observed high-resolution spectra with a Gaussian profile of increasing widths.\\

The method described in this paper makes use of the spectroscopic diagnoses measured precisely on high-resolution spectra. For that, the spectrograph should reproduced as closely as possible the line profile. This is the case of stabilised, fiber-fed spectrographs like SOPHIE and HARPS. The various diagnoses could be measured simultaneously with the radial velocity with no extra observational cost. The detection of line-profile variations at the early stage of the observation could actually avoid wasting telescope time observing false positives rather than planets. However, with slit-fed spectrographs, such as HiReS@Keck, the shape of the line-profile is correlated with the slit illumination profile, which strongly limits the measurement of precision spectroscopic diagnoses \citep{2013ApJ...770..119W}. The use of the iodine cell allows one to monitor the instrumental profile needed to measure precision radial velocities \citep{1996PASP..108..500B}, but limits the determination of the line profile shape. In such case, applying the method described here is impossible, and other observations are needed to firmly rule out blend scenarios, which has an extra observational cost. For these reasons, stabilised fiber-fed spectrographs are more suitable for radial velocity surveys and the follow-up of transiting planet candidates.\\

It was already known that high-resolution spectroscopy can constrain the presence of \textit{resolved} systems down to a few magnitude fainter than the target star \citep[e.g.][]{2012ApJ...745..120B}. We show in this paper, that it is also the case for spectroscopically \textit{unresolved} systems. However, in the latter case, one should however analyse carefully the line profile variation. In the context of validation of small transiting exoplanets, this provides complementary contraints to photometry, which cannot exclude bright contaminants. Some example of constraints provided by high-precision transit photometry can be found in, e.g. \citet{2011ApJ...727...24T, 2015arXiv150101101T, 2011ApJS..197....5F, 2012ApJ...745...81F, 2012Natur.482..195F, 2012ApJ...745..120B, 2013Sci...340..587B}. In all these cases photometric constraints can not reject the presence of a planet transiting a background, chance-aligned star brighter than $\Delta d \approx 6$, which is the domain where spectroscopy is the most sensitive. In the scenario of a background system, it is quite unlikely that such system would be chance-aligned in both the plane of the sky and the RV space. \\

This is not the case for a contaminant physically bounded with the target star, which has a high probability of having a small radial velocity shift. Since the physical companion is most likely blended with the target star, searching for a second set of \textit{resolved} lines in the spectra, as done in the aforementioned papers, is not expected to provide strong constraints to this kind of false-positive scenario. Therefore, without a more careful check of their line-profile variation, as presented in this paper, some of the planets validated in these papers might actually be \textit{larger} planets transiting a star physically-bound with the target. This is still true even if a significant radial velocity variation is detected \citep{2014A&A...571A..37S}.\\

The spectroscopic diagnoses we presented here and the technique developed by \citet{2015AJ....149...18K} can constrain false positive scenarios which are blended in both the plane of the sky and the radial velocity space. Another way to constrain blend contaminant would consist in searching for abnormal stellar lines that do not correspond to the spectral type of the target star. Indeed, if a blended contaminant has a different effective temperature than the target star, there might be some lines (e.g. molecular lines) that are only visible in the spectrum of the coolest component. By checking for extra lines in the spectrum of a target star, one could also constrain the presence of a contaminating star. This can also be done by cross-correlating the observed spectrum with the mask of a M dwarf for which all the lines that are common in both a M-dwarf and G-dwarf masks have been removed. By doing that, if the difference of \teff\, is large enough, the resulting CCF should only present the contribution from the contaminating star.\\

As a benchmark test for the blend model we used in this paper, we analyse with \texttt{PASTIS} the SOPHIE data of the HD16702 system. This system is a blended nearly face-on binary which exhibit a large radial velocity and bisector variations, but no significant FWHM variation. Our model was able to reproduce all the data and to provide constraints on the secondary star that are compatible with the ones derived with the \textit{HIPPARCOS} astrometry \citep{2012A&A...538A.113D} and the technique of \citet{2015AJ....149...18K}.

\section*{Acknowledgments}

We warmly thank the referee, Jon Jenkins, for his thorough and very positive review. We highly appreciate his comments. A.S. is supported by the European Union under a Marie Curie Intra-European Fellowship for Career Development with reference FP7-PEOPLE-2013-IEF, number 627202. AS, NCS, PF acknowledge the support by the European Research Council/European Community under the FP7 through Starting Grant agreement number 239953. NCS also acknowledges the support from Funda\c{c}\~ao para a Ci\^encia e a Tecnologia (FCT) in the form of grant reference PTDC/CTE-AST/098528/2008. PF and NCS acknowledge support by  Funda\c{c}\~ao para a Ci\^encia e a Tecnologia (FCT) through Investigador FCT contracts of reference IF/01037/2013 and IF/00169/2012, respectively, and POPH/FSE (EC) by FEDER funding through the program ``Programa Operacional de Factores de Competitividade - COMPETE''. J.M.A. acknowledges funding from the European Research Council under the ERC Grant Agreement n. 337591-ExTrA.

\appendix

\section[]{The spectroscopic diagnoses}
\label{RVdiag}

\subsection{Radial velocity, FWHM and contrast}

The line profile ($\Psi$) of a dwarf star with a low \vsini can be described by a Gaussian profile (at the spectral resolution of RV-dedicated spectrographs, i.e. between 40 000 and 120 000):
\begin{equation}
\label{Gauss}
\mathcal{G}\left(x\right) = c_{0}\left[1 - c\exp\left(-\frac{1}{2}\frac{\left(x - x_{0}\right)^{2}}{\sigma^{2}}\right)\right],
\end{equation}
where $x$ is the radial velocity, $c_{0}$ is the flux in the continuum, $c$ is the contrast of the line, $x_{0}$ is the radial velocity of the line, and $\sigma$ represents the width of the line. For fast-rotating stars, the line profile can be modeled following the procedure described in \citet{2012A&A...544L..12S} and references therein. For Gaussian profiles, the FWHM is related to $\sigma$ by:
\begin{equation}
\mathrm{FWHM} = \sigma \times2\sqrt{2\ln2} \approx \sigma \times 2.3548
\end{equation}
By fitting the line profile by a Gaussian function, one can therefore measure the radial velocity, the contrast and the FWHM of the stellar line profile.
 
 \begin{figure}
\centering
\includegraphics[width = \columnwidth]{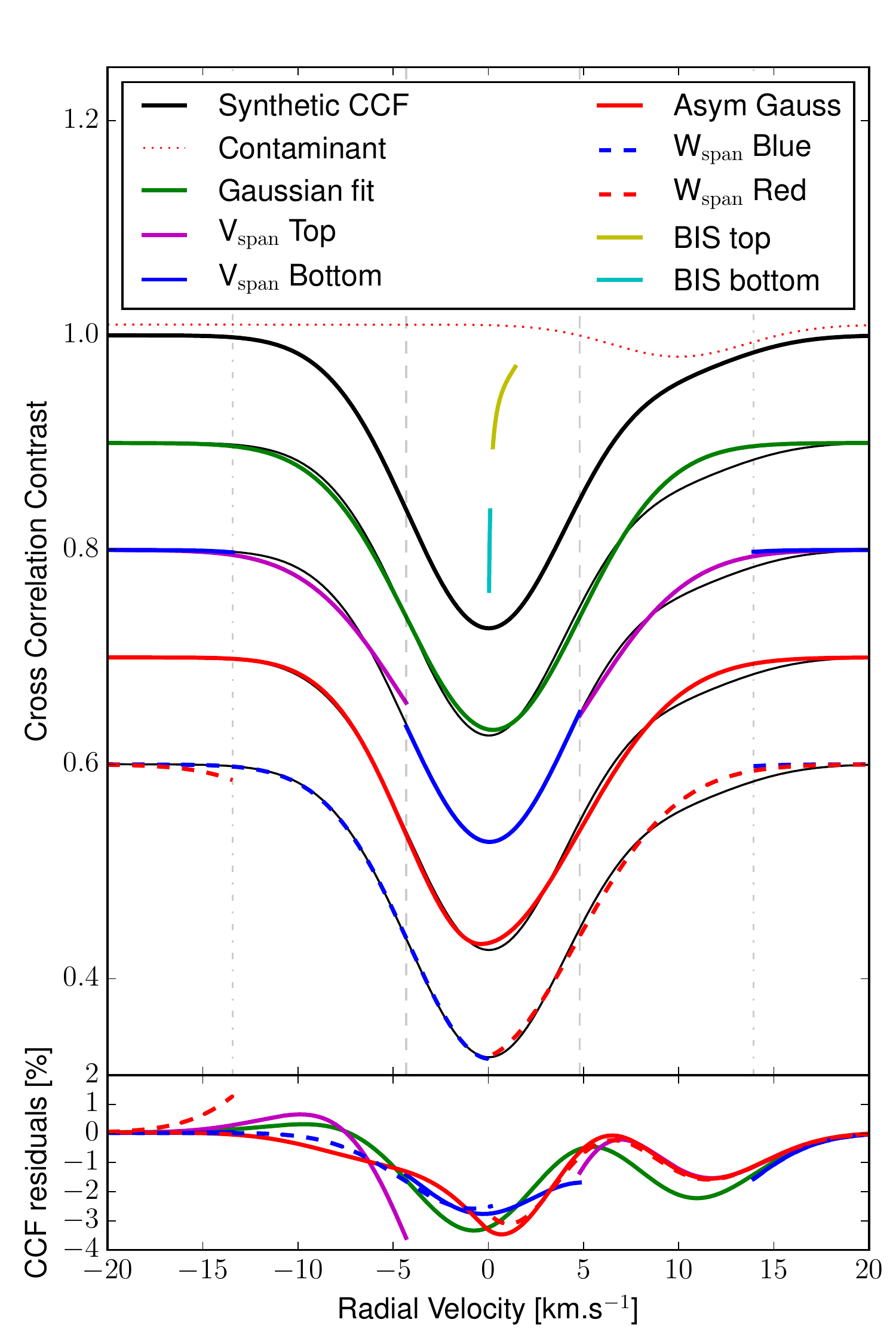}
\caption{\textit{(Upper panel)} Synthetic line profile (thick and thin black lines) contaminated by a diluted star (dotted-red line) with a line contrast of 10\% of the target star, equal FWHM and shifted by 10\kms. The green line displays the Gaussian fit (eq. \ref{Gauss}) to the blended line profile, the magenta line displays the gaussian fit to the top of the line profile (eq. \ref{CCFtop}) while the blue line displays the Gaussian fit to the bottom of the line profile (eq. \ref{CCFbottom}). The solid red line displays the asymmetric Gaussian fit as defined in eq. \ref{AsymGauss}. Cyan and yellow lines display the bisector within the limit defined to compute the bottom and the top parts (respectively) of the BIS. The dashed blue and red lines display the Gaussian fit to the blue (eq. \ref{CCFblue}) and red (eq. \ref{CCFred}) wings of the line profile. The vertical grey dashed and dot-dashed lines represent the $\pm 1\sigma$ and $\pm 3\sigma$ of the line profile, as measured by the symmetric Gaussian fit. \textit{(Lower panel)} Residuals of the fit of the synthetic blended line profile. Displayed lines correspond to the residuals of the fits defined for the upper panel.}
\label{DiagnFig}%
\end{figure}

 \subsection{BIS}
 
To monitor the asymmetry of the line profile several diagnoses have been defined. \citet{1988ApJ...334.1008T} defined the bisector as ``the midpoints of horizontal line segments bounded by the sides of the line profile''. This series of midpoints is not convenient to interpret numerically for each observation. Some diagnoses were therefore suggested in order to describe the asymmetry of the line profile in a single value. This was the case of the bisector curvature, suggested by \citet{1997Natur.385..795G} which consists in computing the ``velocity span of the top portion minus the velocity span of the bottom portion of the bisector'' \citep{1997ApJ...490..412G}. Since this method considers only one measurement in the top and bottom parts of the bisector, it is very sensitive to the noise present in the line profile. To improve this line asymmetry diagnosis, \citet{2001A&A...379..279Q} used the Bisector Inverse Slope (BIS) defined as the velocity span between the average of the top and bottom parts of the bisector:
\begin{equation}
\label{BIS}
\mathrm{BIS} = \mathrm{BIS}_\mathrm{top} - \mathrm{BIS}_\mathrm{bottom},
\end{equation}
where $\mathrm{BIS}_\mathrm{top}$ is the average of the bisector between 60\% and 90\% of the total contrast of the line profile and $\mathrm{BIS}_\mathrm{bottom}$ is the average of the bisector between 10\% and 40\% of the total contrast of the line profile, where 100\% corresponds to the continuum  and 0\% to the minimum of the line profile (see Fig. \ref{DiagnFig}). We find that several different limits have been used in previous studies for the top part of the BIS: 55\% -- 85\% according to \citet{2001A&A...379..279Q}, 55\% -- 90\% according to \citet{2006A&A...454..341D} and 60\% -- 90\% according to \citet{2013A&A...557A..93F}. Since different BIS definitions will lead to different results, a unique definition of how to compute the BIS should be defined to avoid wrong interpretation when comparing BIS data. We propose here to keep the values 60\% -- 90\% as definitive values for BIS computation since they are symmetric with the bottom values around the half maximum of the line.

\subsection{V$_\mathrm{span}$}

To improve the precision of the bisector, \citet{2011A&A...528A...4B} suggested the V$_\mathrm{span}$ diagnosis. Instead of computing the average midpoint of the horizontal line segment which can be noisy for low-SNR spectra, the idea is to compute the radial velocity difference measured by fitting two Gaussian functions, one to the top and one to the bottom parts of the line profile. The Gaussian fit is expected here to be less sensitive to noise in the stellar line than the traditional bisector. The limits between the top and the bottom of the line profile are defined as the $\pm 1 \sigma$-limit from the measured radial velocity ($x_{0}$). The two sub-line profiles ($\Psi_\mathrm{top}$ and $\Psi_\mathrm{bottom}$) are thus defined as (see Fig. \ref{DiagnFig}):
\begin{eqnarray}
\label{CCFtop}
\Psi_\mathrm{top} &=& \Psi\left(\left[- \infty, x_{0}-1\sigma\right] \cup \left[x_{0}+1\sigma, +\infty\right]\right)\\
\label{CCFbottom}
\Psi_\mathrm{bottom} &=& \Psi\left(\left[- \infty, x_{0}-3\sigma\right] \cup \left[x_{0} - 1\sigma, x_{0} + 1\sigma\right]\right.\nonumber\\
&& \cup~\left.\left[x_{0}+3\sigma, +\infty\right]\right)
\end{eqnarray}
The diagnosis V$_\mathrm{span}$ is therefore defined as:
\begin{equation}
\mathrm{V}_\mathrm{span} = x_{0_{top}} - x_{0_{bottom}},
\end{equation}
where $x_{0_{top}}$ and $x_{0_{bottom}}$ are the radial velocities measured to the top of the line profile ($\Psi_\mathrm{top}$, eq. \ref{CCFtop}) and to the bottom of the line profile ($\Psi_\mathrm{bottom}$, eq. \ref{CCFbottom}), respectively.

 \subsection{BiGauss}

\citet{2006A&A...453..309N} defined another diagnosis (called BiGauss) to monitor line asymmetry of pulsating stars. It consists in fitting an asymmetric Gaussian function to the line profile (see Fig. \ref{DiagnFig}):
\begin{equation}
\label{AsymGauss}
\mathcal{G}_\mathrm{asy}\left(x\right) =\begin{cases}
    c_{0}\left[1 - c\exp\left(-\frac{1}{2}\frac{\left(x - x_{0}\right)^{2}}{\left(\sigma\left(1-A\right)\right)^{2}}\right)\right], &  \mathrm{where}\ x < x_{0}\\
    c_{0}\left[1 - c\exp\left(-\frac{1}{2}\frac{\left(x - x_{0}\right)^{2}}{\left(\sigma\left(1+A\right)\right)^{2}}\right)\right], &  \mathrm{where}\ x \geq x_{0}.
\end{cases}
\end{equation}

where $A$ is the percentage of asymmetry (relatively to $\sigma$). The BiGauss diagnosis is defined as the difference between the radial velocity measured with a symmetric Gaussian function (eq. \ref{Gauss}) and the asymmetric BiGaussian function (eq. \ref{AsymGauss}), as a proxy of the line asymmetry:
\begin{equation}
\label{BiGauss}
\mathrm{BiGauss} = x_{0}|_{A=0} - x_{0}|_{A\neq 0}
\end{equation}

 \subsection{V$_\mathrm{asy}$}

A fourth diagnosis was defined recently by \citet{2013A&A...557A..93F} to monitor asymmetry in line profiles caused by stellar activity. This diagnosis, called V$_\mathrm{asy}$, consists in computing the differential spectral information, as defined by \citet{2001A&A...374..733B}, between the red and the blue wings of the line profile:
\begin{equation}
\mathrm{V}_\mathrm{asy} = \frac{\sum_{i}\left(W_{i}^{red} - W_{i}^{blue}\right)\times\overline{W_{i}}}{\sum_{i}\overline{W_{i}}},
\end{equation}
where $W_{i}$ are the weight of the point computed at the flux level $i$ in the line profile \citep[eq. 8 of][]{2001A&A...374..733B}:
\begin{equation}
\label{Wi}
W_{i} = \frac{\lambda_{i}^{2}\left(\frac{\partial \Psi_{i}}{\partial\lambda_{i}}\right)^{2}}{\Psi_{i} + \sigma_{D}^{2}},
\end{equation}
where $\lambda_{i}$ and $\Psi_{i}$ are the wavelength and flux of the spectra, respectively, and $\sigma_{D}^{2}$ the detector readout noise. $W_{i}^{blue}$ and $W_{i}^{red}$ are the blue and red line profile wing information (respectively), and $\overline{W_{i}}$ their average. If computed in the radial velocity space on the CCF, the equation \ref{Wi} needs to be changed to (Figueira et al., in prep.):
\begin{equation}
\label{WiCCF}
W_{i}^{CCF} = \frac{c^{2}\left(\frac{\partial CCF_{i}}{\partial RV_{i}}\right)^{2}}{CCF_{i} + \sigma_{D}^{2}},
\end{equation}
where $RV_{i}$ and $CCF_{i}$ are the radial velocity and the flux in the CCF, respectively, and $c$ is the speed of light.

 \subsection{W$_\mathrm{span}$}

Four methods have been developed so far to estimate the asymmetry of a spectral line, because it's a key diagnosis to correct radial velocities from stellar activity \citep{2011A&A...528A...4B}, from blended contaminants \citep{2002A&A...392..215S, 2012A&A...538A.113D}, and from instrumental effects \citep{2010A&A...523A..88B}. Therefore, more sensitive is the diagnosis to asymmetry, better the correction or the constraints are expected. For that propose, we define here a fifth diagnosis to monitor line asymmetry. It consists on fitting two Gaussian functions to the blue and red wings of the line profile (see Fig. \ref{DiagnFig}): 
\begin{eqnarray}
\label{CCFblue}
\Psi_\mathrm{blue} &=& \Psi\left(\left[- \infty, x_{0}\right] \cup \left[x_{0}+3\sigma, +\infty\right]\right)\\
\label{CCFred}
\Psi_\mathrm{red} &=& \Psi\left(\left[- \infty, x_{0}-3\sigma\right] \cup \left[x_{0}, +\infty\right]\right)
\end{eqnarray}
We can thus defined the asymmetry measurement of this fifth diagnosis, that we call W$_\mathrm{span}$:
\begin{equation}
\mathrm{W}_\mathrm{span} = x_{0_{blue}} - x_{0_{red}},
\end{equation}
where $x_{0_{blue}}$ and $x_{0_{red}}$ are the radial velocity derived from the fit of the blue ($\Psi_\mathrm{blue}$, eq. \ref{CCFblue}) and red  ($\Psi_\mathrm{red}$, eq. \ref{CCFred}) wings of the line profile, respectively.\\

In Fig. \ref{DiagnFig}, we illustrate these five diagnoses by modeling the line profile of a star, with a FWHM of 10\kms\, and a contrast of 30\%. We contaminated it with a star with a line FWHM of 10\kms, a contrast of 3\%, and shifted in radial velocity by +10\kms\ from the main star. This example was chosen to present a huge line-asymmetry, that can be seen by eye. 

\section[]{CCF Modeling}
\label{CCFModel}

To simulate CCFs, we use the equations (B.1), (B.2), (B.3), (B.4) and (B.5) from \citet{2010A&A...523A..88B} for SOPHIE. We first compute the FWHM of the Gaussian function from the \vsini of the star following the equation: 

\begin{equation}
\label{FWHM}
FWHM = \sqrt{\frac{\vsini^{2} + A^{2}\sigma_{0}^{2}}{A^{2}}}\cdot 2\sqrt{2\ln(2)}
\end{equation}

where A is a scaling factor listed in Table \ref{sigma0values}. $\sigma_{0}$ is related to the color index $(B-V)$ in the range 0.4 -- 1.3, similar to FGK dwarfs: 

\begin{equation}
\label{sigma0}
\sigma_{0} = \alpha - \beta (B-V) + \gamma (B-V)^{2} - \delta(B-V)^{3},
\end{equation}
where the coefficient $\alpha$, $\beta$, $\gamma$, and $\delta$ are given in Table \ref{sigma0values}. We then compute the expected contrast $\mathcal{C}$ of the CCF : 

\begin{equation}
\label{contrast}
\mathcal{C} = \frac{\mathcal{W}}{\sqrt{2\pi}}\cdot \frac{2\sqrt{2\ln(2)}}{FWHM},
\end{equation}
where $\mathcal{W}$ is the area of the CCF, expressed as function of the metallicity \met\, and the color index $(B-V)$ : 

\begin{eqnarray}
\label{CCFareaS}
\log\mathcal{W} &=& \frac{\met + a_{S} - b_{S}(B-V)}{c_{S}}\\ \nonumber
\end{eqnarray}
where the coefficient $a_{S}$, $b_{S}$, $c_{S}$ are listed in Table \ref{CCFareaValues}. These relations were calibrated for $0.43 \leq (B - V) \leq 0.98$ 
and for $-0.47 \leq \met \leq 0.44$ for SOPHIE.\\

\begin{table}
  \centering 
  \caption{Coefficients of equations \ref{FWHM} \& \ref{sigma0}}
  \label{sigma0values}
  \setlength{\tabcolsep}{1.mm}
  \begin{tabular}{c|ccccc}
instru. mode & $\alpha$ & $\beta$ & $\gamma$ & $\delta$ & A \\
\hline
SOPHIE -- HE & 10.52 & 22.56 & 22.37 & 6.95 & 1.64 \\
SOPHIE -- HR & 9.90 & 22.56 & 22.37 & 6.95 & 1.73 \\
\end{tabular}
\end{table}

\begin{table}
  \centering 
  \caption{Coefficients of equations \ref{CCFareaS}.}
  \label{CCFareaValues}
  \setlength{\tabcolsep}{3mm}
  \begin{tabular}{c|ccc}
\multicolumn{4}{c}{SOPHIE}\\
 & $a_{S}$ & $b_{S}$ & $c_{S}$\\
\hline   
G2 &  -0.9440 &  1.4992 & 3.8807 \\
K5 & 0.2615  &   2.2606 & 3.9553 \\
 & & & \\
\end{tabular}
\end{table}

The line profile are assumed to have a Gaussian profile. We therefore compute the average line profile for each star considered ($CCF_{i}$) as follow.
\begin{equation}
\label{CCFeq}
CCF_{i}\left(x\right) = \mathcal{F}_{i}\left[1 - \mathcal{C}_{i}\cdot\exp\left(-\frac{\left(x - x_{0}\right)^{2}}{2\cdot\left(\frac{FWHM_{i}}{2\sqrt{2\ln(2)}}\right)^{2}}\right)\right]
\end{equation}

where $\mathcal{F}_{i}$ is the flux of star in the continuum integrated within the spectrograph bandpass that we assumed to be equivalent to the Johnson-V band for SOPHIE. The blended synthetic CCF is therefore: 
\begin{equation}
\label{CCFBlend}
CCF \left(x\right) = \frac{\sum_{i}CCF_{i}\left(x\right)}{\sum_{i}\mathcal{F}_{i}}
\end{equation}

We then use this synthetic CCF to measure the various spectroscopic diagnoses by fitting it with a Levenberg -- Marquardt algorithm as implemented into the Scipy package (optimize.leastsq).

\section[]{Computation time and numerical precision and accuracy of the models}
\label{ComputTime}

Measuring the diagnoses to a large sample of observations, or through a large MCMC analysis takes time. The computation speed as well as the precision and accuracy of the fitting procedure should be part of the equation when defining the best diagnosis to use. These elements should also depend on the number of data points sampling the line profile. To quantify them, we simulated a blended Gaussian line profile. We defined a target line profile with a FWHM of 10 \kms, a contrast of 30\% and a contaminating line profile with a FWHM of 12 \kms, a contrast of 30\%, a flux ratio of 1/100 compared with the target star, and a constant radial velocity shift of 10\kms. We used different sampling rate for the simulation of this blended line profile, ranging from 10\cms\, to 1\kms. Then, for each sampling size, we simulated 100 times the same system, by shifting the blended line profile randomly within a uniform distribution with a width of 1\kms. By doing that, the blended line profile was sampled differently in all the simulations. On these synthetic line profile we measured the various spectroscopic diagnoses aforementioned as well as the time taken to compute them using the optimize.leastsq module embedded in the scipy package in python2.7. We used the default values for this function, except for the xtol parameter that we sed to 10$^{-5}$. This experiment was done with an Intel Xeon CPU cadenced at 2.27GHz. Note that the evaluation of the symmetric and asymmetric Gaussian functions (used to derive the radial velocity, the FWHM, BiGauss, \vspan, and \wspan) is implemented within \texttt{PASTIS} in a C++ module, together with their Jacobian functions, in order to speed up the fitting procedure. \\

\begin{figure}
\begin{center}
\includegraphics[width=\columnwidth]{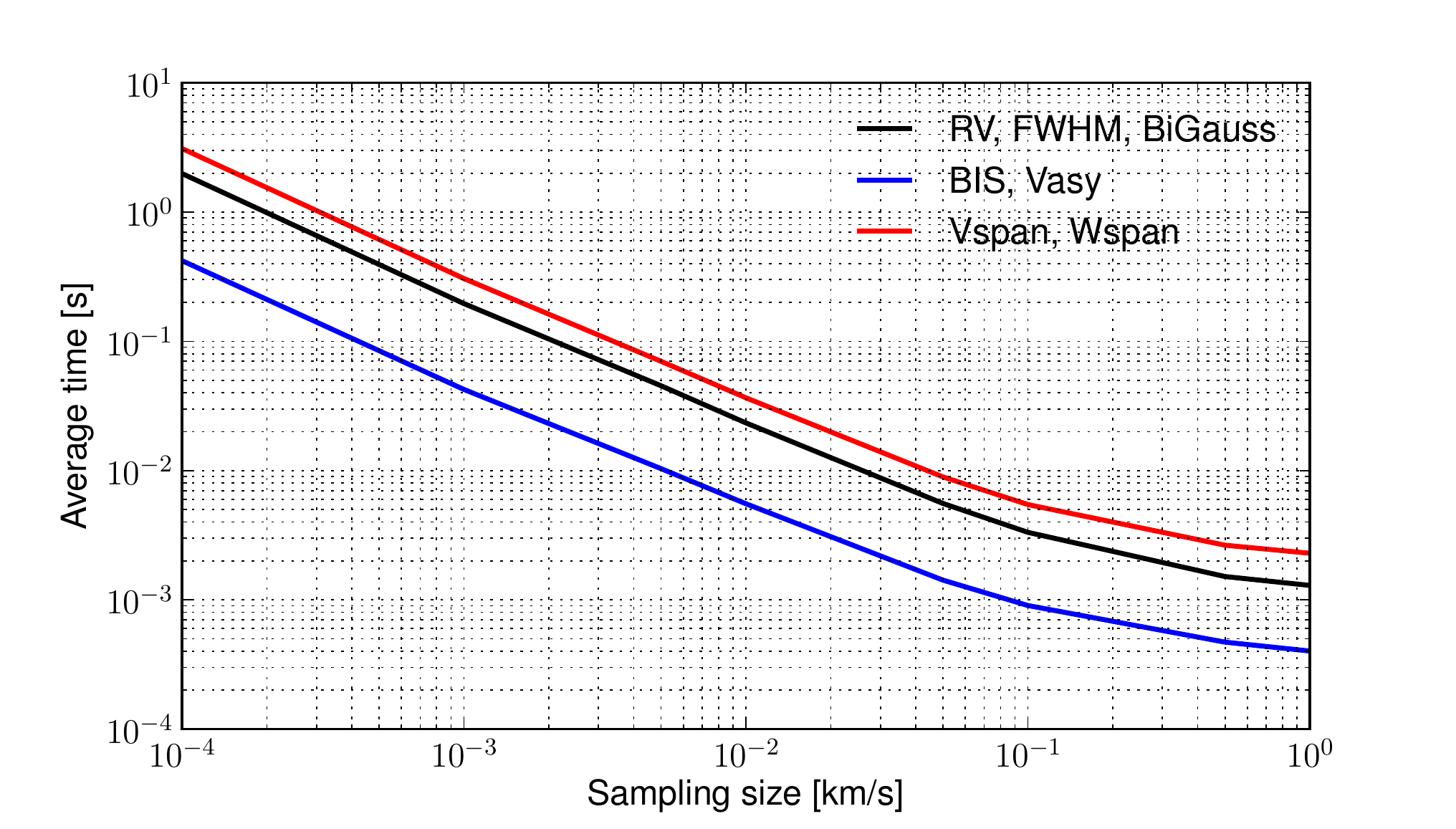}\\
\includegraphics[width=\columnwidth]{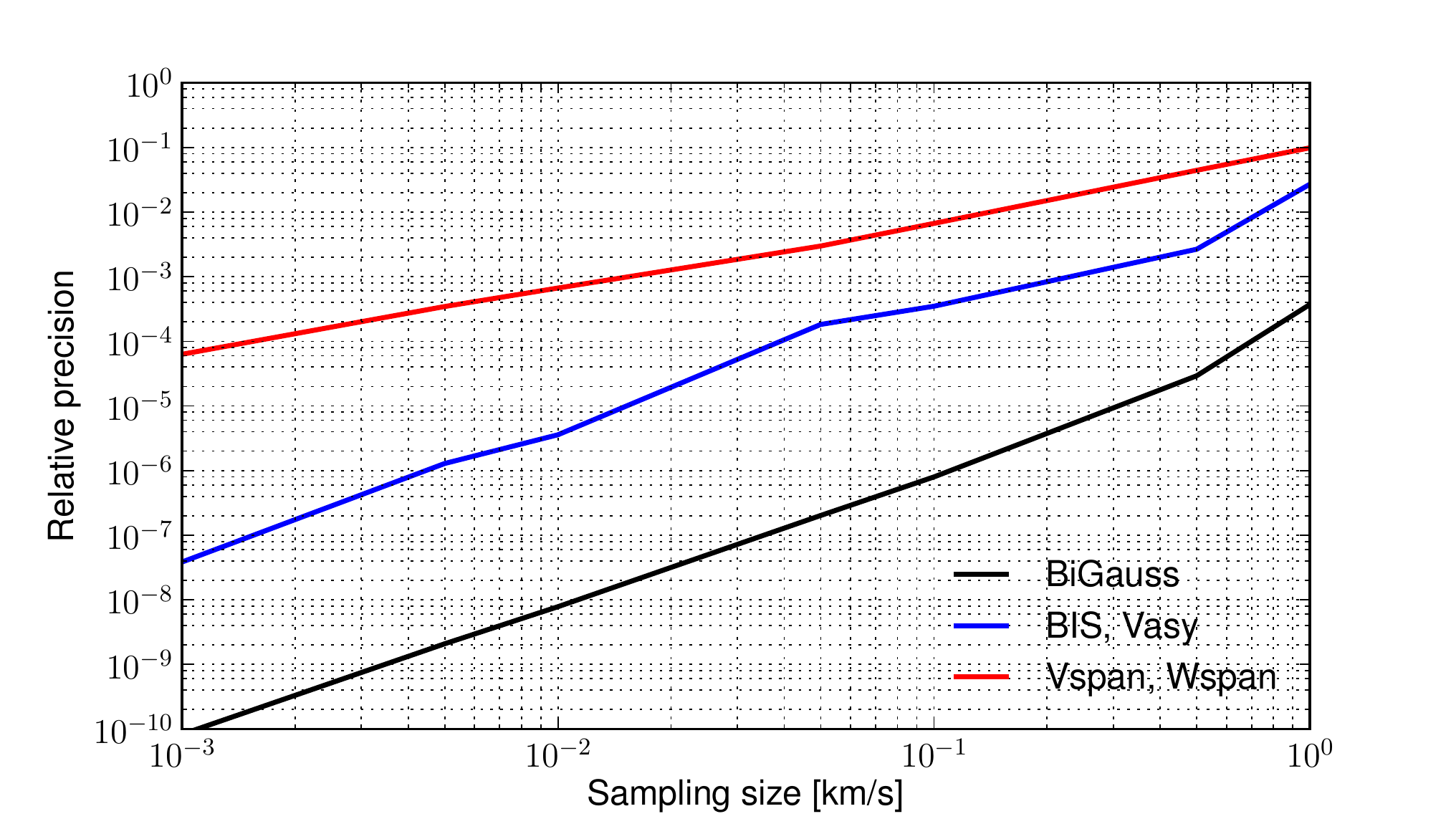}\\
\includegraphics[width=\columnwidth]{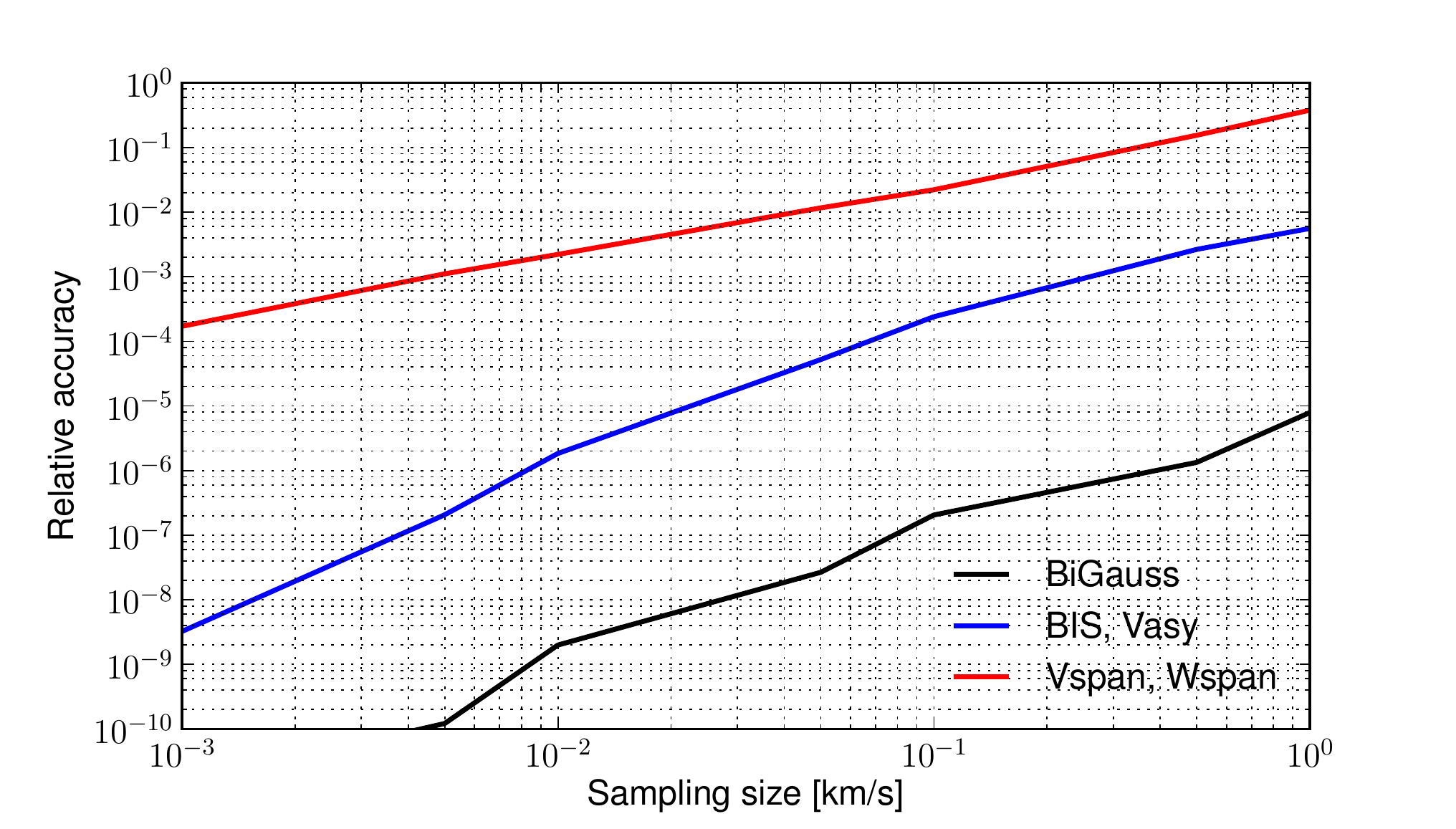}
\caption{\textit{Top panel:} Average time to measure the different spectroscopic diagnoses on a noiseless line profile as function of the sampling size (see text for details). \textit{Middle and bottom panels:} Relative precision (middle panel) and accuracy (bottom panel) of the asymmetry diagnoses on a noiseless blended line profile as function of the sampling size. This assumes a line profile with a FWHM of 10\kms. The RV and FWHM are not displayed here since they are at the level of the numerical noise.}
\label{ModelTime}
\end{center}
\end{figure}

Figure \ref{ModelTime} displays the mean time to compute the various diagnoses as function of the sampling size. Note that this simulation has been performed with a line profile sampled from -50\kms\, to +50\kms\, around the target star. Decreasing or increasing the wavelength or radial velocity range is expected to scale down or up (respectively) these results. As expected, the computation time increase exponentially when decreasing the sampling size. The diagnoses BIS and \vasy\, which does not require any minimisation algorithms are the fastest ones. The diagnoses \vspan\, and \wspan\, which require two fitting procedure, take twice more time than the measurement of the radial velocity, FWHM, and BiGauss, as expected.\\

The precision and accuracy of the fitting model should also depend on the number of points in the line profile. Thus, we also investigated the relative precision and accuracy of our models as function of the sampling size. We estimated the precision of the diagnoses by computing the RMS of the measurements for the 100 simulations aforementioned. The accuracy has been estimated by comparing the median value of the 100 simulations at different sampling size with the one with a sampling size of 10\cms. We therefore assumed here that the 10\cms\, sampling has a negligible noise, compared with the other simulations. We then normalised the precision and accuracy with the median value of the asymmetry found for a sampling size of 10\cms, to derive the relative precision and accuracy. Figure \ref{ModelTime} displays the relative precision and accuracy to measure the various asymmetry diagnoses as function of the sampling of the line profile. We find that BiGauss is the most precise and accurate diagnoses to measure even with relatively a poor sampling of the line profile. On the other hand, \vspan\, and \wspan\, need a good sampling of the line profile to be measured precisely and accurately. This behaviour could be explained by the fact that the BiGauss diagnosis consists in fitting a continuous function, while \vspan\, and \wspan\, consist in fitting two discontinuous functions. BIS and \vasy\, follow the same precision and accuracy trend, and are dominated by the numerical noise produced by the interpolation of the line profile. The sampling of the line profile needs to be increased by a factor of more than 10 for BIS and \vasy\, to reach the same precision and accuracy as BiGauss. Given the computation time, the relative precision and accuracy, the sensitivity of the diagnoses to blends (Fig. \ref{AllDiagn}) and to noise (see Section. \ref{photnoise}), BiGauss is the most efficient spectroscopic diagnosis to constrain blended contamination.\\

\section[]{Table of the parameters for the \texttt{PASTIS} analysis of the system HD16702}
\label{Anal16702}

\begin{table*}
\caption{Free parameters of the HD16702 analysis with their prior and posterior distributions. For the posterior we report only the median value and its 68.3\% uncertainty. $\mathcal{U}(a,b)$ represents a Uniform prior between $a$ and $b$; $\mathcal{N}(\mu,\sigma^{2})$ represents a Normal distribution with a mean of $\mu$ and a width of $\sigma^{2}$; $\mathcal{N_{U}}(\mu,\sigma^{2}, a, b)$ represents a Normal distribution with a mean of $\mu$ and a width of $\sigma^{2}$ and limited by a Uniform distribution between $a$ and $b$.}
\begin{center}
\begin{tabular}{lcc}
\hline
\hline
Parameter & Prior & Posterior \\
\hline
\multicolumn{3}{l}{\textit{Primary star}}\\
Effective temperature \teff\, [K] & $\mathcal{N}(5908;25)$ & $5909 \pm 26$\\
Surface gravity \logg\, [g.cm$^{-2}$] & $\mathcal{N}(4.46; 0.03)$ & $4.45 \pm 0.03$\\
Iron abundance \met\, [dex] & $\mathcal{N}(-0.12;0.03)$ & $-0.12 \pm 0.03$\\
Equatorial velocity \vsini$_{1}$ [\kms] & $\mathcal{N_{U}}(2.56;1;0;20)$ & $3.42 \pm 0.04$\\
Systemic radial velocity $\gamma$ [\kms] & $\mathcal{U}(0;10)$ & $4.576 \pm 0.004$\\
\hline
\multicolumn{3}{l}{\textit{Secondary star}}\\
Initial mass M$_{2}$ [\Msun] & $\mathcal{N_{U}}(0.55;0.14;0.1;20)$ & $0.35 \pm 0.03$\\
Equatorial velocity \vsini$_{2}$ [\kms] & $\mathcal{U}(0.5;15)$ & $6.4 \pm 2.1$\\
\hline
\multicolumn{3}{l}{\textit{Binary orbit}}\\
Orbital period $P$ [d] & $\mathcal{N}(72.8322;0.2)$ & $72.839 \pm 0.003$\\
Periastron epoch $T_{p}$ [BJD - 2450000] & $\mathcal{N}(4983.45;12)$ & $4983.20 \pm 0.14$\\
Eccentricity $e$ & $\mathcal{U}(0;1)$ & $0.138 \pm 0.002$\\
Argument of periastron $\omega$ [\degr] & $\mathcal{U}(0;360)$ & $256.0 \pm 0.7$\\
Inclination $i_{orb}$ [\degr] & $\mathcal{U}(0;90)$ & $9.4 \pm 0.5$\\
\hline
\multicolumn{3}{l}{\textit{Data-related parameters}}\\
Radial velocity offset G2 -- K5 $\Delta_{RV}$ [\ms] & $\mathcal{U}(-200;200)$ & $-1 \pm 5$\\
Radial velocity jitter G2 $\sigma_{j_{RV_{G2}}}$ [\ms] & $\mathcal{U}(0;1000)$ & $15 \pm 3$\\
Radial velocity jitter K5 $\sigma_{j_{RV_{K5}}}$ [\ms] & $\mathcal{U}(0;1000)$ & $15 \pm 3$\\
BiGauss offset G2 $\gamma_{BG_{G2}}$ [\ms] & $\mathcal{U}(-200;200)$ & $-1 \pm 3$\\
BiGauss offset K5 $\gamma_{BG_{K5}}$ [\ms] & $\mathcal{U}(-200;200)$ & $-71 \pm 3$\\
BiGauss jitter G2 $\sigma_{j_{BG_{G2}}}$ [\ms] & $\mathcal{U}(0;1000)$ & $3 \pm 3$\\
BiGauss jitter K5 $\sigma_{j_{BG_{K5}}}$ [\ms] & $\mathcal{U}(0;1000)$ & $4 \pm 4$\\
FWHM offset G2 -- K5 $\Delta_{FW}$ [\ms] & $\mathcal{U}(-200;200)$ & $7 \pm 6$\\
FWHM jitter G2 $\sigma_{j_{FW_{G2}}}$ [\ms] & $\mathcal{U}(0;1000)$ & $8 \pm 6$\\
FWHM jitter K5 $\sigma_{j_{FW_{K5}}}$ [\ms] & $\mathcal{U}(0;1000)$ & $9 \pm 6$\\
\hline
\hline
\end{tabular}
\end{center}
\label{params16702}
\end{table*}%

\bsp

\label{lastpage}


\begin{thebibliography}{99}
\bibitem[Allard et al.(2012)]{2012IAUS..282..235A} Allard, F., Homeier, D., \& Freytag, B.\ 2012, IAU Symposium, 282, 235 
\bibitem[Almenara et al.(2009)]{2009A&A...506..337A} Almenara, J.~M., Deeg, H.~J., Aigrain, S., et al.\ 2009, \aap, 506, 337
\bibitem[Anglada-Escud{\'e} \& Butler(2012)]{2012ApJS..200...15A} Anglada-Escud{\'e}, G., \& Butler, R.~P.\ 2012, \apjs, 200, 15
\bibitem[Baranne et al.(1996)]{1996A&AS..119..373B} Baranne, A., Queloz, D., Mayor, M., et al.\ 1996, \aaps, 119, 373
\bibitem[Barclay et al.(2013)]{2013ApJ...768..101B} Barclay, T., Burke, C.~J., Howell, S.~B., et al.\ 2013, \apj, 768, 101
\bibitem[Batalha et al.(2010)]{2010ApJ...713L.103B} Batalha, N.~M., Rowe, J.~F., Gilliland, R.~L., et al.\ 2010, \apjl, 713, L103 
\bibitem[Batalha et al.(2013)]{2013ApJS..204...24B} Batalha, N.~M., Rowe, J.~F., Bryson, S.~T., et al.\ 2013, \apjs, 204, 24
\bibitem[Boisse et al. (2010)]{2010A&A...523A..88B} Boisse, I., Eggenberger, A., Santos, N.~C., et al.\ 2010, \aap, 523, A88
\bibitem[Boisse et al.(2011a)]{2011A&A...528A...4B} Boisse, I., Bouchy, F., H{\'e}brard, G., et al.\ 2011a, \aap, 528, A4
\bibitem[Boisse et al.(2011b)]{2011EPJWC..1602003B} Boisse, I., Bouchy, F., Chazelas, B., et al.\ 2011b, European Physical Journal Web of Conferences, 16, 02003
\bibitem[Boisse et al.(2012)]{2012A&A...545A.109B} Boisse, I., Bonfils, X., \& Santos, N.~C.\ 2012, \aap, 545, AA109
\bibitem[Borucki et al.(2012)]{2012ApJ...745..120B} Borucki, W.~J., Koch, D.~G., Batalha, N., et al.\ 2012, \apj, 745, 120
\bibitem[Borucki et al.(2013)]{2013Sci...340..587B} Borucki, W.~J., Agol, E., Fressin, F., et al.\ 2013, Science, 340, 587
\bibitem[Bouchy et al.(2001)]{2001A&A...374..733B} Bouchy, F., Pepe, F., \& Queloz, D.\ 2001, \aap, 374, 733
\bibitem[Bouchy et al.(2009a)]{2009IAUS..253..129B} Bouchy, F., Moutou, C., Queloz, D., \& the \textit{CoRoT} Exoplanet Science Team 2009a, IAU Symposium, 253, 129, arXiv:0902.3520 
\bibitem[Bouchy et al.(2009b)]{2009A&A...505..853B} Bouchy, F., H\'ebrard, G., Udry, S., et al.\ 2009b, \aap, 505, 853
\bibitem[Bouchy et al.(2013)]{2013A&A...549A..49B} Bouchy, F., D{\'{\i}}az, R.~F., H{\'e}brard, G., et al.\ 2013, \aap, 549, A49
\bibitem[Bou{\'e} et al.(2012)]{2012MNRAS.422L..57B} Bou{\'e}, G., Oshagh, M., Montalto, M., \& Santos, N.~C.\ 2012, \mnras, 422, L57
\bibitem[Bou{\'e} et al.(2013)]{2013A&A...550A..53B} Bou{\'e}, G., Montalto, M., Boisse, I., Oshagh, M., \& Santos, N.~C.\ 2013, \aap, 550, A53
\bibitem[Brown et al.(2014)]{2014arXiv1412.7761B} Brown, D.~J.~A., Anderson, D.~R., Armstrong, D.~J., et al.\ 2014, arXiv:1412.7761
\bibitem[Buchhave et al.(2011)]{2011ApJS..197....3B} Buchhave, L.~A., Latham, D.~W., Carter, J.~A., et al.\ 2011, \apjs, 197, 3
\bibitem[Butler et al.(1996)]{1996PASP..108..500B} Butler, R.~P., Marcy, G.~W., Williams, E., et al.\ 1996, \pasp, 108, 500
\bibitem[Cabrera et al.(2009)]{2009A&A...506..501C} Cabrera, J., Fridlund, M., Ollivier, M., et al.\ 2009, \aap, 506, 501
\bibitem[Cameron(2012)]{2012Natur.492...48C} Cameron, A.~C.\ 2012, \nat, 492, 48 
\bibitem[Carone et al.(2012)]{2012A&A...538A.112C} Carone, L., Gandolfi, D., Cabrera, J., et al.\ 2012, \aap, 538, A112
\bibitem[Cavarroc et al.(2012)]{2012Ap&SS.337..511C} Cavarroc, C., Moutou, C., Gandolfi, D., et al.\ 2012, \apss, 337, 511
\bibitem[Col{\'o}n et al.(2012)]{2012MNRAS.426..342C} Col{\'o}n, K.~D., Ford, E.~B., \& Morehead, R.~C.\ 2012, \mnras, 426, 342
\bibitem[Coughlin et al.(2014)]{2014AJ....147..119C} Coughlin, J.~L., Thompson, S.~E., Bryson, S.~T., et al.\ 2014, \aj, 147, 119
\bibitem[Cunha et al.(2013)]{2013A&A...550A..75C} Cunha, D., Figueira, P., Santos, N.~C., Lovis, C., \& Bou{\'e}, G.\ 2013, \aap, 550, A75
\bibitem[Dall et al.(2006)]{2006A&A...454..341D} Dall, T.~H., Santos, N.~C., Arentoft, T., Bedding, T.~R., \& Kjeldsen, H.\ 2006, \aap, 454, 341
\bibitem[Deeg et al.(2009)]{2009A&A...506..343D} Deeg, H.~J., Gillon, M., Shporer, A., et al.\ 2009, \aap, 506, 343 
\bibitem[Delfosse et al.(2013)]{2013sf2a.conf..497D} Delfosse, X., Donati, J.-F., Kouach, D., et al.\ 2013, SF2A-2013: Proceedings of the Annual meeting of the French Society of Astronomy and Astrophysics, 497
\bibitem[D{\'{\i}}az et al.(2012)]{2012A&A...538A.113D} D{\'{\i}}az, R.~F., Santerne, A., Sahlmann, J., et al.\ 2012, \aap, 538, A113
\bibitem[D{\'{\i}}az et al.(2014)]{2014MNRAS.441..983D} D{\'{\i}}az, R.~F., Almenara, J.~M., Santerne, A., et al.\ 2014, \mnras, 441, 983
\bibitem[Dotter et al.(2008)]{2008ApJS..178...89D} Dotter, A., Chaboyer, B., Jevremovi{\'c}, D., et al.\ 2008, \apjs, 178, 89
\bibitem[Dumusque et al.(2012)]{2012Natur.491..207D} Dumusque, X., Pepe, F., Lovis, C., et al.\ 2012, \nat, 491, 207
\bibitem[Dumusque et al.(2014)]{2014ApJ...796..132D} Dumusque, X., Boisse, I., \& Santos, N.~C.\ 2014, \apj, 796, 132
\bibitem[Eggenberger \& Udry(2010)]{2010EAS....41...27E} Eggenberger, A., \& Udry, S.\ 2010, EAS Publications Series, 41, 27
\bibitem[Erikson et al.(2012)]{2012A&A...539A..14E} Erikson, A., Santerne, A., Renner, S., et al.\ 2012, \aap, 539, A14
\bibitem[Figueira et al.(2013)]{2013A&A...557A..93F} Figueira, P., Santos, N.~C., Pepe, F., Lovis, C., \& Nardetto, N.\ 2013, \aap, 557, AA93
\bibitem[Fressin et al.(2011)]{2011ApJS..197....5F} Fressin, F., Torres, G., D{\'e}sert, J.-M., et al.\ 2011, \apjs, 197, 5
\bibitem[Fressin et al.(2012a)]{2012ApJ...745...81F} Fressin, F., Torres, G., Pont, F., et al.\ 2012a, \apj, 745, 81
\bibitem[Fressin et al.(2012b)]{2012Natur.482..195F} Fressin, F., Torres, G., Rowe, J.~F., et al.\ 2012b, \nat, 482, 195
\bibitem[Fressin et al.(2013)]{2013ApJ...766...81F} Fressin, F., Torres, G., Charbonneau, D., et al.\ 2013, \apj, 766, 81
\bibitem[Galassi et al. (2011)]{Galassi11} Galassi, M., Davies, J., Theiler, J., Gough, B., Jungman, G., Booth, M., \& Rossi, F., 2011, GNU Scientific Library - Reference manual, Version 1.15, Sec. 21.7.
\bibitem[Gray(1997)]{1997Natur.385..795G} Gray, D.~F.\ 1997, \nat, 385, 795
\bibitem[Gray \& Hatzes(1997)]{1997ApJ...490..412G} Gray, D.~F., \& Hatzes, A.~P.\ 1997, \apj, 490, 412
\bibitem[Hatzes \& Cochran(1992)]{1992ESOC...40..275H} Hatzes, A.~P., \& Cochran, W.~D.\ 1992, European Southern Observatory Conference and Workshop Proceedings, 40, 275
\bibitem[Holman et al.(2010)]{2010Sci...330...51H} Holman, M.~J., Fabrycky, D.~C., Ragozzine, D., et al.\ 2010, Science, 330, 51
\bibitem[Howard et al.(2012)]{2012ApJS..201...15H} Howard, A.~W., Marcy, G.~W., Bryson, S.~T., et al.\ 2012, \apjs, 201, 15 
\bibitem[Hwang et al.(2013)]{2013ApJ...778...55H} Hwang, K.-H., Choi, J.-Y., Bond, I.~A., et al.\ 2013, \apj, 778, 55
\bibitem[Kolbl et al.(2015)]{2015AJ....149...18K} Kolbl, R., Marcy, G.~W., Isaacson, H., \& Howard, A.~W.\ 2015, \aj, 149, 18
\bibitem[Konacki et al.(2003)]{2003ApJ...597.1076K} Konacki, M., Torres, G., Sasselov, D.~D., \& Jha, S.\ 2003, \apj, 597, 1076
\bibitem[Lillo-Box et al.(2014)]{2014A&A...566A.103L} Lillo-Box, J., Barrado, D., \& Bouy, H.\ 2014, \aap, 566, AA103
\bibitem[Lissauer et al.(2012)]{2012ApJ...750..112L} Lissauer, J.~J., Marcy, G.~W., Rowe, J.~F., et al.\ 2012, \apj, 750, 112
\bibitem[Lissauer et al.(2014)]{2014ApJ...784...44L} Lissauer, J.~J., Marcy, G.~W., Bryson, S.~T., et al.\ 2014, \apj, 784, 44
\bibitem[Marcy et al.(2014)]{2014ApJS..210...20M} Marcy, G.~W., Isaacson, H., Howard, A.~W., et al.\ 2014, \apjs, 210, 20
\bibitem[Mayor et al.(2003)]{2003Msngr.114...20M} Mayor, M., Pepe, F., Queloz, D., et al.\ 2003, The Messenger, 114, 20
\bibitem[Mazeh \& Faigler(2010)]{2010A&A...521L..59M} Mazeh, T., \& Faigler, S.\ 2010, \aap, 521, L59
\bibitem[Mordasini et al.(2009)]{2009A&A...501.1161M} Mordasini, C., Alibert, Y., Benz, W., \& Naef, D.\ 2009, \aap, 501, 1161
\bibitem[Morton \& Johnson(2011)]{2011ApJ...738..170M} Morton, T.~D., \& Johnson, J.~A.\ 2011, \apj, 738, 170
\bibitem[Morton(2012)]{2012ApJ...761....6M} Morton, T.~D.\ 2012, \apj, 761, 6
\bibitem[Moutou et al.(2009)]{2009A&A...506..321M} Moutou, C., Pont, F., Bouchy, F., et al.\ 2009, \aap, 506, 321 
\bibitem[Nardetto et al.(2006)]{2006A&A...453..309N} Nardetto, N., Mourard, D., Kervella, P., et al.\ 2006, \aap, 453, 309
\bibitem[Pepe et al.(2002)]{2002A&A...388..632P} Pepe, F., Mayor, M., Galland, F., et al. \ 2002, \aap, 388, 632
\bibitem[Prusti(2012)]{2012AN....333..453P} Prusti, T.\ 2012, Astronomische Nachrichten, 333, 453
\bibitem[Queloz et al.(2001)]{2001A&A...379..279Q} Queloz, D., Henry, G. W., Sivan, J. P., et al.\ 2001, \aap, 379, 279 
\bibitem[Queloz et al.(2009)]{2009A&A...506..303Q} Queloz, D., Bouchy, F., Moutou, C., et al.\ 2009, \aap, 506, 303
\bibitem[Quirrenbach et al.(2014)]{2014SPIE.9147E..1FQ} Quirrenbach, A., Amado, P.~J., Caballero, J.~A., et al.\ 2014, \procspie, 9147, 91471F
\bibitem[Perruchot et al.(2008)]{2008SPIE.7014E..17P} Perruchot, S., Kohler, D., Bouchy, F., et al.\ 2008, \procspie, 7014
\bibitem[Perruchot et al.(2011)]{2011SPIE.8151E..15P} Perruchot, S., Bouchy, F., Chazelas, B., et al.\ 2011, \procspie, 8151, 815115
\bibitem[Petigura et al.(2013)]{2013PNAS..11019273P} Petigura, E.~A., Howard, A.~W., \& Marcy, G.~W.\ 2013, Proceedings of the National Academy of Science, 110, 19273
\bibitem[Rauer et al.(2014)]{2014ExA....38..249R} Rauer, H., Catala, C., Aerts, C., et al.\ 2014, Experimental Astronomy, 38, 249
\bibitem[Ricker et al.(2014)]{2014SPIE.9143E..20R} Ricker, G.~R., Winn, J.~N., Vanderspek, R., et al.\ 2014, \procspie, 9143, 914320
\bibitem[Rowe et al.(2014)]{2014ApJ...784...45R} Rowe, J.~F., Bryson, S.~T., Marcy, G.~W., et al.\ 2014, \apj, 784, 45
\bibitem[Santerne et al.(2011)]{2011EPJWC..1102001S} Santerne, A., Endl, M., Hatzes, A., et al.\ 2011, European Physical Journal Web of Conferences, 11, 02001
\bibitem[Santerne et al.(2012a)]{2012A&A...544L..12S} Santerne, A., Moutou, C., Barros, S.~C.~C., et al.\ 2012a, \aap, 544, L12
\bibitem[Santerne et al.(2012b)]{2012A&A...545A..76S} Santerne, A., D{\'{\i}}az, R.~F., Moutou, C., et al.\ 2012b, \aap, 545, A76
\bibitem[Santerne et al.(2013a)]{2013A&A...557A.139S} Santerne, A., Fressin, F., D{\'{\i}}az, R.~F., et al.\ 2013a, \aap, 557, AA139
\bibitem[Santerne et al.(2013b)]{2013sf2a.conf..555S} Santerne, A., D{\'{\i}}az, R.~F., Almenara, J.-M., et al.\ 2013b, SF2A-2013: Proceedings of the Annual meeting of the French Society of Astronomy and Astrophysics, 555
\bibitem[Santerne et al.(2014)]{2014A&A...571A..37S} Santerne, A., H{\'e}brard, G., Deleuil, M., et al.\ 2014, \aap, 571, AA37
\bibitem[Santos et al.(2002)]{2002A&A...392..215S} Santos, N.~C., Mayor, M., Naef, D., et al.\ 2002, \aap, 392, 215
\bibitem[Santos et al.(2014)]{2014A&A...566A..35S} Santos, N.~C., Mortier, A., Faria, J.~P., et al.\ 2014, \aap, 566, AA35
\bibitem[Shporer et al.(2011)]{2011AJ....142..195S} Shporer, A., Jenkins, J.~M., Rowe, J.~F., et al.\ 2011, \aj, 142, 195
\bibitem[Toner \& Gray(1988)]{1988ApJ...334.1008T} Toner, C.~G., \& Gray, D.~F.\ 1988, \apj, 334, 1008
\bibitem[Torres et al.(2004)]{2004ApJ...614..979T} Torres, G., Konacki, M., Sasselov, D.~D., \& Jha, S.\ 2004, \apj, 614, 979
\bibitem[Torres et al.(2005)]{2005ApJ...619..558T} Torres, G., Konacki, M., Sasselov, D.~D., \& Jha, S.\ 2005, \apj, 619, 558
\bibitem[Torres et al.(2011)]{2011ApJ...727...24T} Torres, G., Fressin, F., Batalha, N.~M., et al.\ 2011, \apj, 727, 24
\bibitem[Torres et al.(2015)]{2015arXiv150101101T} Torres, G., Kipping, D.~M., Fressin, F., et al.\ 2015, \apj, in press, arXiv:1501.01101
\bibitem[Wright et al.(2013)]{2013ApJ...770..119W} Wright, J.~T., Roy, A., Mahadevan, S., et al.\ 2013, \apj, 770, 119
\bibitem[Zucker \& Mazeh(1994)]{1994ApJ...420..806Z} Zucker, S., \& Mazeh, T.\ 1994, \apj, 420, 806
\bibitem[Zurlo et al.(2013)]{2013A&A...554A..21Z} Zurlo, A., Vigan, A., Hagelberg, J., et al.\ 2013, \aap, 554, AA21
\end{thebibliography}
\end{document}